\documentclass[11pt]{article}
\usepackage{jheppub}
\usepackage{amsmath,amssymb,amsfonts,graphicx,slashed,color, mathtools, amsthm}
\usepackage{comment}
\usepackage{enumerate}
\usepackage{float}
\usepackage{caption}
\usepackage{subcaption}
\usepackage{pdflscape}
\usepackage{physics}
\usepackage{tikz}
\usepackage{mathrsfs}
\usepackage{dsfont}
\usetikzlibrary{arrows.meta, decorations.markings}
\tikzset{
    pil/.style={->, thick, shorten <=2pt, shorten >=2pt,},
    mid arrow/.style={postaction={decorate,decoration={
        markings,
        mark=at position .5 with {\arrow[#1]{stealth'}}
    }}},
}

\newtheorem{theorem}{Theorem}

\newtheorem{proposition}{Proposition}
\newtheorem{definition}{Definition}
\newtheorem{lemma}{Lemma}

\newtheorem{innercustomthm}{Theorem}
\newenvironment{mythm}[1]
  {\renewcommand\theinnercustomthm{#1}\innercustomthm}
  {\endinnercustomthm}

\newtheorem{innercustomlemma}{Lemma}
\newenvironment{mylemma}[1]
  {\renewcommand\theinnercustomlemma{#1}\innercustomlemma}
  {\endinnercustomlemma}

\renewcommand{\tr}{{\rm tr}}

\title{Black holes as black boxes: a query lower bound on python's lunch decoding}

\preprint{CERN-TH-2026-298}

\author[1, 2]{Chris Waddell}
\author[3, 4]{Michelle Xu}

\affiliation[1]{Perimeter Institute for Theoretical Physics, 31 Caroline St N, Waterloo, ON\ N2L 2Y5, Canada}
\affiliation[2]{Theoretical Physics Department, CERN, CH-1211 Geneva 23, Switzerland}
\affiliation[3]{Simons Institute for the Theory of Computing, UC Berkeley, Melvin Calvin Laboratory, 2190, Berkeley CA 94720, USA}
\affiliation[4]{Joint Center for Quantum Information and Computer Science (QuICS), University of Maryland, 3100 Atlantic Building, College Park MD 20742-2420, USA}

\emailAdd{christopher.neil.h.waddell@cern.ch}

\abstract{The python’s lunch conjecture relates the complexity of bulk reconstruction in holographic tensor networks to an exponential cost associated with post-selection. We establish a tight query lower bound for this decoding problem in a Haar random oracle model, with a python's lunch circuit that applies Haar random unitary \(U\) to a logical input and fixed ancillas, followed by post-selection on \(m\) output qubits. Given one copy of the normalized encoded state and coherent oracle access to \(U\) and \(U^\dagger\), including controlled queries, we prove that recovering even a single logical bit with constant average advantage requires \(\Omega(2^{m/2})\) queries. Together with the state-independent amplitude amplification upper bound for approximately isometric encodings, this establishes query complexity \(\Theta(2^{m/2})\) at constant decoding accuracy. We also comment on the decoding cost for other ``admissible'' ensembles of unitaries $U$ which give rise to families of approximately isometric holographic maps.}

\begin{document}

\maketitle

\section{Introduction} \label{sec:intro}

Operational questions in quantum gravity are constrained not only by what is permitted by the laws of quantum mechanics, but also by what an observer can feasibly compute. This observation has played an increasingly important role in attempts to reconcile the semi-classical description of black holes with their underlying unitary dynamics. A foundational example is the work of Harlow and Hayden \cite{Harlow:2013tf}, who argued that decoding the correlations between early Hawking radiation and a late Hawking quantum requires a computation exponential in the black hole entropy. The decoding experiment underlying the firewall paradox may therefore be information-theoretically possible but computationally impossible before the black hole evaporates. In this way, computational complexity can enforce a form of black hole complementarity: semi-classical effective field theory need only remain consistent against experiments that a computationally limited observer can actually perform.

\begin{figure}
    \centering
    \includegraphics[width=0.9\textwidth]{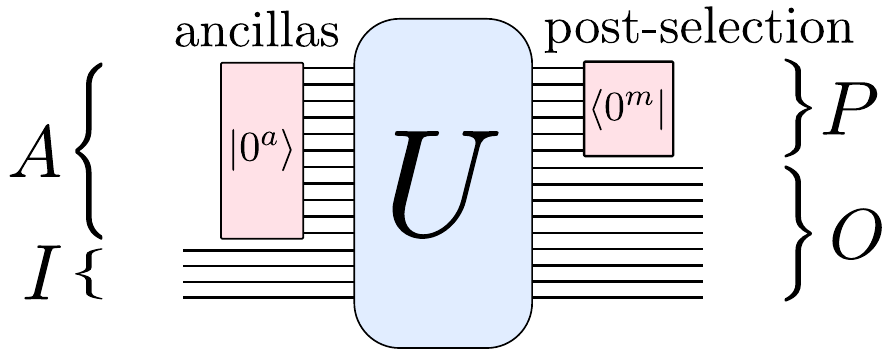}
    \caption{General form of holographic maps considered by Brown et al. in \cite{Brown:2019rox}, to which the python's lunch conjecture is expected to hold for suitable $U$. The map is from input register $I$ to output register $O$, with ancillas and post-selection on $A$ and $P$ respectively.  
    }
    \label{fig:map}
\end{figure}

A more geometric manifestation of this idea is the python's lunch conjecture (PLC) of Brown, Gharibyan, Penington, and Susskind \cite{Brown:2019rox}, whose motivation comes from tensor network models of holography. Consider a tensor network, interpreted as a bulk-to-boundary map $K_{U} : \mathcal{H}_{I} \rightarrow \mathcal{H}_{O}$, which 
can be written as an efficiently implementable unitary \(U\), with \(a\) ancilla qubits initialized in 
the zero state and \(m\) output qubits post-selected onto 
the zero state,
\begin{equation}
K_U
=
\bigl(\langle 0^m|_P\otimes \mathds{1}_O\bigr)
U
\bigl(\mathds{1}_I\otimes |0^a\rangle_A\bigr) \: .
\label{eq:intro_KU}
\end{equation}
See figure \ref{fig:map}. 
The resulting map is a sensible encoding of the logical input state, provided that it is approximately proportional to an isometry,
\begin{equation}
2^m K_U^\dagger K_U \approx \mathds{1}_I \: .
\label{eq:intro_isometry}
\end{equation}
Under this assumption, information is not destroyed by the projection and can in principle be recovered from the normalized state
\begin{equation}
|\Psi(\psi,U)\rangle
=
\frac{K_U|\psi\rangle}
{\| K_U|\psi\rangle \|_2} \: .
\label{eq:intro_postselected_state}
\end{equation}

However, such a recovery faces a computational obstruction. A direct attempt to implement the holographic map via a ``measure-and-pray'' protocol for the post-selection succeeds with probability $O(2^{-m})$, and therefore requires $O(2^{m})$ attempts to succeed with probability $O(1)$. 
Improving on this, Brown et al. showed that a state-independent version of amplitude amplification yields a unitary implementation of the holographic map using $O(2^{m/2})$ applications of \(U\) and \(U^\dagger\), together with polynomially many elementary operations per application. 
Inverting this unitary gives a bulk reconstruction protocol, provided our state is promised to be in the code subspace. 
The python's lunch conjecture asserts, roughly, that when the tensor network is sufficiently generic and contains no additional structure that can be exploited by the decoder, this upper bound is optimal up to polynomial factors.

In gravity, the corresponding obstruction is described by a collection of quantum extremal surfaces. For a boundary region whose entanglement wedge contains a python's lunch, the globally minimal quantum extremal surface $X_{\text{RT}}$ is separated from the boundary by another, locally minimal extremal surface $X_{\text{app}}$, often called the \emph{appetizer}. Between them lies a bulge surface $X_{\text{bulge}}$ obtained by an appropriate ``maximinimax'' prescription 
\cite{Brown:2019rox, Engelhardt:2023bpv, Arora:2024edk}. The python's lunch conjecture assigns exponentially large complexity to the reconstruction of operators lying behind the appetizer,
\begin{equation}
\mathcal C_{\text{PL}}
\sim
\mathcal C_0
\exp\left[
\frac{1}{2}
\left(
S_{\mathrm{gen}}(X_{\mathrm{bulge}})
-
S_{\mathrm{gen}}(X_{\mathrm{app}})
\right)
\right] \: ,
\label{eq:intro_PL_gravity}
\end{equation}
where \(\mathcal C_0\) is expected to be polynomial in the semi-classical parameters. In a regime where bulk-entropy corrections are subleading, this becomes
\begin{equation}
\mathcal C_{\text{PL}}
\sim
\operatorname{poly}(1/G_N)
\exp\left[
\frac{
A(X_{\mathrm{bulge}})
-
A(X_{\mathrm{app}})
}{8G_N}
\right] \: .
\label{eq:intro_PL_area}
\end{equation}
For an evaporating black hole, the region behind the appetizer contains precisely the interior information whose recovery from the Hawking radiation is obstructed in the Harlow-Hayden experiment. The python's lunch proposal therefore gives a geometric explanation for the Harlow-Hayden complexity barrier.

There is some indirect evidence for this proposal in gravity, in the form of a converse. Building on 
work of Engelhardt and Wall \cite{Engelhardt:2017aux, Engelhardt:2018kcs} (and of course the work of HKLL \cite{Hamilton:2006az}), Engelhardt, Penington, and Shahbazi-Moghaddam showed \cite{Engelhardt:2021mue} that operators in the exterior of the outermost extremal surface can be reconstructed using simple boundary sources, Lorentzian timefolds, and an HKLL-type reconstruction procedure. Thus, the outermost extremal surface is an obstruction to this class of simple reconstruction protocols. Other work has supplied indirect tests of the projective tensor network model; in particular, cryptographic features of the model lead to 
boundary correlation inequalities that can be checked geometrically in AdS/CFT \cite{May:2024epy}. 
These results support aspects of the proposed relationship between geometry and complexity, but they are far from establishing the conjectured circuit complexity lower bound for the reconstruction task itself.

Even in the tensor network setting, such a lower bound is difficult to prove, as is ``par for the course'' in complexity theory. In \cite{Brown:2019rox}, Brown et al. pointed out a general complexity-theoretic reason to not expect an efficient procedure for implementing arbitrary rare post-selections. 
The possibility that exponentially unlikely post-selected computations could always be simulated in quantum polynomial time is tantamount to the statement
\begin{equation}
\mathsf{BQP}=\mathsf{PostBQP} \: .
\end{equation}
Since \(\mathsf{PostBQP}=\mathsf{PP}\) \cite{Aaronson:2004voc}, where $\mathsf{PP}$ is a very powerful class (which for example includes $\mathsf{NP}$), this would imply the implausible collapse \(\mathsf{BQP}=\mathsf{PP}\). This argument rules strongly against the possibility of eliminating the exponential cost of post-selection altogether. It does not, however, show that the particular square-root scaling \(2^{m/2}\) is optimal for the reconstruction problem of interest: it remains compatible, for example, with a hypothetical decoder of complexity \(2^{m/3}\). 

Faced with the familiar difficulty of proving circuit lower bounds, our work will make the standard strategic retreat to the black-box setting. We regard \(U\) as an unknown unitary oracle and allow a decoder to make coherent queries to \(U\) and \(U^\dagger\) (and controlled versions thereof), while all computation independent of \(U\) is free. We then ask how many queries are required to invert the normalized post-selected map.

It is sufficient to study a particularly simple restriction of the full reconstruction problem. Considering the bulk state with a single unknown bit
\begin{equation}
| e_k (n) \rangle
\equiv
|k\rangle|0\rangle^{\otimes(n-1)} \: ,
\qquad k\in \{0,1\} \: ,
\end{equation}
we define the corresponding boundary state by
\begin{equation}
|\Psi_k(U)\rangle
\equiv 
\frac{\Pi U |e_k (n) \rangle}
{\| \Pi U | e_k (n) \rangle \|_2} \: , \qquad \Pi = |0^m\rangle\langle0^m|\otimes\mathds{1}_{2^{n-m}} \: ,
\label{eq:intro_one_bit_states}
\end{equation}
for $U$ an $n$-qubit unitary. 
Our decoder is given one copy of \(|\Psi_k(U)\rangle\), where \(k\) is uniformly random, together with oracle access to \(U\) and \(U^\dagger\), and must determine \(k\). 
Our main result is a matching query lower bound for Haar random \(U\).

\medskip
\noindent
\textbf{Main theorem (informal).}
\emph{Let \(U\) be Haar random on \(n\) qubits, let \(1\leq m<n\), and let \(\mathcal A\) be a quantum algorithm which receives one copy of \(|\Psi_k(U)\rangle\) and makes at most \(T\) queries to \(U\) and \(U^\dagger\), including controlled queries. Its average probability of correctly identifying \(k\) satisfies}
\begin{equation}
\mathbb E_{U}
\bigl[p_{\mathrm{succ}}(U,\mathcal A)\bigr]
\leq \frac12 + 6\sqrt{2}\,T\,2^{-m/2} \: .
\label{eq:intro_main_bound}
\end{equation}
\emph{Consequently, average success probability at least \(2/3\) requires}
\begin{equation}
T
\geq
\frac{1}{36\sqrt{2}} 2^{m/2} \: .
\label{eq:intro_main_corollary}
\end{equation}
Together with the amplitude amplification upper bound, this establishes query complexity
\begin{equation}
Q_{\mathrm{PL}}=\Theta(2^{m/2})
\end{equation}
for constant accuracy reconstruction in the Haar random black box model.

The black box qualification is essential to the interpretation of this result. In the 
typical formulation of a bulk reconstruction task, the microscopic description of the holographic map is known, and a decoder may inspect that description and exploit its algebraic or geometric structure. In our model, by contrast, all dependence on \(U\) must enter through applications of \(U\) and \(U^\dagger\). A query lower bound therefore does not imply a circuit lower bound for a known, explicitly specified unitary. It instead establishes that no \emph{black box} or \emph{relativizing} decoder can improve on the square-root post-selection penalty: any improvement must use details of the structure of \(U\).\footnote{A related query complexity result is the work of Yang and Engelhardt on learning Haar random unitary dynamics \cite{Yang:2023zic}. Their learner must use its query access to the \emph{dynamics} to construct a hypothesis for the global action of \(U\) before receiving a new test state, after which it no longer has oracle access. Our decoder instead receives a particular \(U\)-dependent post-selected state, retains query access to a piece of the \emph{holographic map} throughout the computation, and need only recover one logical bit. Their unitary-learning lower bound therefore does not imply ours. To our knowledge, no previous work has established a matching query lower bound for the python’s lunch decoding problem.}

This distinction is especially important because Haar random unitaries do not
themselves provide realistic models of efficiently implementable holographic
maps.  Moreover, 
alternative notions of randomness are not
sufficient to ensure hard decoding.  For example, we note that taking $U$ to be a uniformly random Clifford
unitary gives rise, with high probability, to an exact stabilizer isometry $K_{U}$ with
a polynomial-size decoder.  
More generally, every post-selected stabilizer map that is
proportional to an isometry admits an efficient Clifford decoder \cite{Cleve:1996pp}. 
This latter observation applies also to holographic stabilizer tensor networks,
not only to the non-geometric ensemble of global random Cliffords. Random
stabilizer tensor networks themselves can define bulk-to-boundary isometries while
reproducing the Ryu-Takayanagi formula \cite{Ryu:2006bv} and entanglement wedge reconstruction \cite{Czech:2012bh, Headrick:2014cta, Wall:2012uf};
some constructions also exhibit complementary recovery, and map local bulk
Hamiltonians to local boundary Hamiltonians
\cite{Pastawski:2015qua, Hayden:2016cfa, Nezami:2016zni, Apel:2021tnn}.
Nevertheless, whenever their bulk-to-boundary map is a stabilizer isometry, it
can be inverted on the code subspace by a polynomial-size stabilizer circuit. 
Of course, stabilizer networks
remain highly idealized models of AdS/CFT, and for connected,
non-stabilizer tensor networks of polynomial depth, we do not presently know how to prove an exponential query or circuit lower bound.  Our result should
therefore be viewed as isolating the post-selection obstruction in a
structureless oracle model, and as a first step toward understanding which additional properties of holographic 
tensor networks make this obstruction persist. 

The remainder of the paper is organized as follows. In section \ref{sec:pld-problem} we formulate the python's lunch decoding problem, introduce an admissibility condition for ensembles of holographic maps ensuring information-theoretic recoverability, and review the amplitude amplification upper bound. 
We also discuss the dependence of the conjecture on the ensemble of
holographic maps, recalling that post-selected stabilizer maps admit efficient decoders.   
In section \ref{sec:lower-bound} we prove the Haar random query lower bound, 
and we conclude in section \ref{sec:discussion} with limitations and possible extensions of the result.

\textbf{Note added.} After the main result of this work and most of this manuscript were completed, the work \cite{Cochran:2026vha} appeared. This paper considers a Haar random oracle access version of the black hole decoding problem of Harlow-Hayden \cite{Harlow:2013tf}, which is closely related to our Haar random python's lunch decoding problem, though our proof techniques establishing theorem \ref{thm:query-lower-bound} substantially differ from theirs.

\subsection{Notation and conventions}
\label{subsec:conventions}

All vector spaces in this paper are finite-dimensional Hilbert
spaces.  For a vector $x$, we write
\begin{equation}
    \|x\|_2 \equiv \sqrt{\langle x,x\rangle} \: .
\end{equation}
For a linear map $M:\mathcal H_{1} \to \mathcal H_{2}$, the notation
\begin{equation}
    \|M\|
    \equiv
    \sup_{\|x\|_2=1}\|Mx\|_2
\end{equation}
denotes the operator norm induced by the Euclidean norm.  Equivalently, the operator norm is given by
\begin{equation}
    \|M\|
    =
    \sqrt{\lambda_{\max}(M^\dagger M)} \: ,
\end{equation}
with $\lambda_{\max}(\cdot)$ the maximum eigenvalue, 
so $\|M\|$ is the largest singular value of \(M\).  If \(M\) is
Hermitian, then this is the largest eigenvalue in absolute value. We will make use of the triangle inequality and sub-multiplicativity, 
\begin{equation}
    \|AB\|\leq \|A\|\,\|B\| \: .
\end{equation}
We will also occasionally use the Frobenius norm,
\begin{equation}
    \| M \|_{F} \equiv \sqrt{\text{tr}(M^{\dagger} M)} \: ,
\end{equation}
which satisfies $\| \cdot \| \leq \| \cdot \|_{F}$.

\section{The python's lunch decoding problem}
\label{sec:pld-problem}

We are interested in the question of the complexity of decoding quantum information inside of a python's lunch. To investigate this, we will consider 
a generalization of the setting proposed by \cite{Brown:2019rox}, wherein the encoding map is modelled by an isometry drawn from some ensemble, followed by post-selection on a subset of qubits. 
We will denote such maps by
\begin{equation} \label{eq:KU}
    K_{U} \equiv (\langle 0^{m} |_{P} \otimes \mathds{1}_{O}) U (\mathds{1}_{I} \otimes |0^{a} \rangle_{A}) \: , 
\end{equation}
with $U$ unitary, $a$ the number of ancilla qubits, and $m$ the number of post-selected qubits. 
We will always use $n = \log_{2}(d)$ to denote the number of qubits on which $U$ acts, so $K_{U}$ maps from $n-a$ to $n-m$ qubits. 
We will use the notation $A$ for the ancilla register, $P$ for the post-selected register, $I$ for the input register of $K_{U}$, and $O$ for the output register. We record our conventions and frequently used notation in tables \ref{tab:registers} and \ref{tab:notation} for reference.

\begin{table}[t]
    \centering
    \begin{tabular}{ccl}
        \hline
        Register & Number of qubits & Description \\
        \hline
        $I$ & $n-a$ & Logical input register \\
        $A$ & $a$ & Ancilla input register, initialized in $\ket{0^a}$ \\
        \hline
        $O$ & $n-m$ & Remaining output register after post-selection \\
        $P$ & $m$ & Output register post-selected onto $\ket{0^m}$ \\
        \hline
        $R$ & -- & Auxiliary/reference system \\
        $\hat{I}$ & $n-a$ & Reconstructed logical register output by the decoder \\
        \hline
    \end{tabular}
    \caption{Register conventions used throughout. The unitary $U$ maps
    $IA$ to $OP$. We also occasionally denote $d = 2^{n}$, $q = 2^{-m}$, and $r = 2^{n-m}$.}
    \label{tab:registers}
\end{table}

\begin{table}[t]
    \centering
    \begingroup
    \renewcommand{\arraystretch}{1.5}
    \begin{tabular}{cc}
        \hline
        Notation & Description \\
        \hline
        $K_{U} \equiv (\langle 0^{m} |_{P} \otimes \mathds{1}_{O}) U (\mathds{1}_{I} \otimes |0^{a} \rangle_{A})$ & Holographic map \\
        $\Pi \equiv |0^{m} \rangle \langle 0^{m}|_{P} \otimes \mathds{1}_{O}$ & Post-selection projector \\
        $J \equiv \mathds{1}_{I} \otimes | 0^{a} \rangle_{A}$ & Ancilla preparation \\
        $| e_{k}(n) \rangle \equiv | k \rangle |0 \rangle^{n - 1}$ & Computational basis states with first bit $k$ \\
        $|\Psi(\psi, U)\rangle \equiv \frac{(K_U\otimes\mathds{1}_R)|\psi\rangle }{\|(K_U\otimes\mathds{1}_R)|\psi\rangle \|_2} $ & Normalized post-selected state \\
        $|\Psi_{k}(U)\rangle \equiv \frac{(K_U\otimes\mathds{1}_R) | e_{k}(n) \rangle }{\|(K_U\otimes\mathds{1}_R) | e_{k}(n) \rangle  \|_2} $ & Normalized post-selected state for $| \psi \rangle = |e_{k}(n) \rangle$
    \end{tabular}
    \endgroup
    \caption{Definitions for frequently used notation.}
    \label{tab:notation}
\end{table}

In fact, we need to specify a bit more structure in order to ensure that such encoding maps are (approximately) isometric with high probability, and that decoding is thus information-theoretically achievable. 
\begin{definition} \label{def:admissible}
    Let $n > a > m > 0$ be integers. We will call an ensemble $\mathcal{E}$ of unitaries $U \in U(2^{n})$ \emph{$(a, m, \eta, \delta)$-admissable} if 
    \begin{equation}
        \underset{U \sim \mathcal{E}}{\textnormal{Pr}} \left[ \| 2^{m} K_{U}^{\dagger} K_{U} - \mathds{1}_{I} \| \leq \eta \right] \geq 1 - \delta \: .
    \end{equation}
\end{definition}

\noindent
As an example, we can see that the Haar ensemble is admissable with suitable parameters. 

\begin{theorem} \label{thm:Haar}
    The Haar ensemble on unitaries $U(2^{n})$ is $(a, m, \eta, \delta)$-admissible whenever $0 < \eta \leq 1$ and
    \begin{equation}
        2^{n - m} \geq C \eta^{-2} \left( 2^{n-a} + c_{1} \ln(1 / \delta) + c_{2} \right) 
    \end{equation}
    for some universal constants $C, c_{1}, c_{2} >0$. 
\end{theorem}

We can provide a proof of this theorem making use of standard results in the theory of random matrices (see e.g. \cite{Meckes2019RandomMatrix}); we defer this to appendix \ref{app:admissibility}. 
Having identified an appropriate class of holographic maps, we may now state the python's lunch decoding problem.

\begin{definition}[Python's Lunch Decoding Problem]
Fix integers $n>a>m>0$ and parameters $\eta,\delta,\zeta, \epsilon>0$, and
let $\mathcal E$ be an $(a,m,\eta,\delta)$-admissible ensemble of
unitaries $U\in U(2^n)$.
For any reference register $R$ and any pure state
$|\psi\rangle_{IR}$ for which the post-selection probability is
non-zero, define the normalized post-selected state
\begin{equation}
    |\Psi(\psi, U)\rangle_{OR}
    \equiv
    \frac{(K_U\otimes\mathds{1}_R)|\psi\rangle_{IR}}
    {\|(K_U\otimes\mathds{1}_R)|\psi\rangle_{IR}\|_2} \: .
\end{equation}
For a decoding channel
\begin{equation}
    \mathcal D_U:
    \mathcal B(\mathcal H_O)
    \longrightarrow
    \mathcal B(\mathcal H_{\widehat I}) \: ,
\end{equation}
define its worst-case decoding error on the instance $U$ by\footnote{This is technically distinct from the diamond norm distance between $(\mathcal{D}_{U} \circ K_{U}) \otimes \mathds{1}_{R}$ and the identity due to the normalization in $| \Psi(\psi, U) \rangle$, though the two agree when $K_{U}$ is an exact isometry.}
\begin{equation}
\label{eq:PL-decoding-error}
    \operatorname{err}(U,\mathcal D_U)
    \equiv
    \sup_{\textnormal{dim}(R), \: |\psi\rangle_{IR}}
    \frac12
    \left\|
    (\mathcal D_U\otimes \mathds{1}_R)
    \bigl(
        |\Psi(\psi, U)\rangle\langle\Psi(\psi, U)|
    \bigr)
    -
    |\psi\rangle\langle\psi|_{\hat{I} R}
    \right\|_1 \: .
\end{equation}
If $K_U$ has a non-trivial kernel, we set
$\operatorname{err}(U,\mathcal D_U)=1$.

A solution to the
\emph{$(n,a,m,\eta,\delta,\zeta, \epsilon, \mathcal E)$-python's lunch decoding problem} is a quantum algorithm implementing channels $\{ \mathcal D_U\}_{U}$
such that
\begin{equation}
    \Pr_{U\sim\mathcal E}
    \left[
        \operatorname{err}(U,\mathcal D_U) \leq \zeta
    \right]
    \geq 1 - \epsilon \: .
\end{equation}
\end{definition}

\noindent
Note that, given our admissibility requirement, the python's lunch decoding problem is by construction information-theoretically solvable for some quantum channel.

In fact, for our purposes, it will be sufficient to restrict to perhaps the simplest case of this decoding problem, where the goal is to reconstruct a single classical bit. 

\begin{definition}[One-Bit Python's Lunch Decoding Problem] Fix integers $n > a > m > 0$ and $\eta, \delta, \zeta, \epsilon > 0$, and let $\mathcal{E}$ be an $(a, m, \eta, \delta)$-admissible ensemble of unitaries $U \in U(2^{n})$. 
Let 
\begin{equation}
    |e_{i}(n) \rangle \equiv | i \rangle | 0 \rangle^{\otimes (n-1)} \: , \qquad i \in \{0, 1\} \: ,
\end{equation}
and define the projection
\begin{equation}
    \Pi \equiv | 0^{m} \rangle \langle 0^{m} | \otimes \mathds{1}_{O} \: , 
\end{equation}
and post-selection probability
\begin{equation}
    p_{k}(U) \equiv \| \Pi U | e_{k} (n) \rangle \|_{2}^{2} \: .
\end{equation}
Whenever $p_{k}(U) > 0$, define the normalized states
\begin{equation}
    | \Psi_{k}(U) \rangle \equiv \frac{\Pi U | e_{k}(n) \rangle}{\sqrt{p_{k}(U)}} \: .
\end{equation}
If either $p_0(U)$ or $p_1(U)$ vanishes, set
$p_{\mathrm{succ}}(U)=0$.

A solution to the \emph{$(n, a, m, \eta, \delta, \zeta, \epsilon, \mathcal{E})$-one-bit python's lunch decoding problem} is a quantum algorithm $\mathcal{A}$ outputting a single classical bit such that
\begin{equation}
    \Pr_{U\sim\mathcal E}
    \left[
        p_{\mathrm{succ}}(U)<1-\zeta
    \right]
    \leq \epsilon \: .
\end{equation}
\end{definition}

\noindent
It is straightforward to see that the task of reconstructing a single classical bit should be equivalent to or easier than reconstructing a more general quantum input in terms of computational complexity.\footnote{Note in particular that $\text{err}(U, \mathcal{D}_{U}) \leq \zeta$ implies that the algorithm $\mathcal{A}$ defined by applying $\mathcal{D}_{U}$ and then measuring the first qubit in the computational basis gives success probability $p_{\text{succ}}(U) \geq 1 - \zeta$ for the one-bit decoding task. } In section \ref{sec:lower-bound}, we will show a query lower bound on the one-bit python's lunch decoding problem, so the same lower bound is immediately applicable to the more general python's lunch decoding problem.

\subsection{Review: upper bound on python's lunch decoding} \label{sec:ub}

Having established the python's lunch decoding problem as an information-theoretically solvable computational problem, a natural question is whether it is possible to characterize the computational complexity of this problem. This is indeed the content of the python's lunch conjecture \cite{Brown:2019rox}. Roughly speaking, this is the statement that for some physically reasonable admissible ensemble of unitaries $\mathcal{E}$, e.g. a suitable class of holographic tensor network unitaries, the python's lunch decoding problem has circuit complexity controlled by the number of post-selections according to
\begin{equation}
    \mathcal{C}_{\text{PL}} = \Theta(2^{m/2}) \: ,
\end{equation}
up to polynomial prefactors.

Two important qualifications are required here. First of all, the result proven in \cite{Brown:2019rox} is in fact only an upper bound $\mathcal{C}_{\text{PL}} = O(2^{m/2})$ on the complexity, while the lower bound remains conjectural (though as mentioned in the introduction, some strong arguments for an exponential lower bound are given). Second of all, there is some ambiguity in which admissible $\mathcal{E}$ this conjecture should apply to; while the upper bound is argued to hold under the assumption that $\mathcal{E}$ is sufficiently scrambling (e.g. a unitary 2-design) and $a \gg m \gg 1$, it is less clear what conditions are sufficient to ensure the desired lower bound. 

In this subsection, we will focus on upper bounds, briefly reviewing the upper bound on $\mathcal{C}_{\text{PL}}$ proven by \cite{Brown:2019rox}, using an approach inspired by Grover's search algorithm \cite{Grover:1996rk}. 
It will turn out that admissibility of $\mathcal{E}$ is a sufficient criterion for the upper bound to be applicable. 
We will state the result and proof somewhat formally, though the key ingredient in the proof is Brown et al.'s state-independent amplification protocol; a more intuitive introduction to this can be found in appendix A of \cite{Brown:2019rox}. 

\begin{theorem}[Brown-Gharibyan-Penington-Susskind {\cite{Brown:2019rox}}]
\label{thm:grover-upper-bound}
Let $\mathcal E$ be an $(a,m,\eta,\delta)$-admissible ensemble, and suppose $\eta\leq \frac{1}{2}$.  Suppose furthermore that $U$ and $U^\dagger$
can be implemented with circuit complexity $\mathcal C(U)$.

Then there exists a universal constant $c>0$ and, for every $U$, a
decoding channel $\mathcal D_U$ of circuit complexity
\begin{equation}
    \mathcal{C}_{\textnormal{PL}} = O\!\left[
        \bigl(\mathcal C(U)+\textnormal{poly}(n)\bigr)2^{m/2}
    \right]
\end{equation}
such that, on the admissibility event $\| 2^{m} K_{U}^{\dagger} K_{U} - \mathds{1}_{I} \| \leq \eta$, we obtain
\begin{equation}
    \operatorname{err}(U,\mathcal D_U)
    \leq
    \frac{\pi}{2} \left(2^{-m/2} + c \eta \right) \: .
\end{equation}
In particular, for every
\begin{equation} \label{eq:zeta_lb}
    \zeta
    \geq
    \frac{\pi}{2} \left(2^{-m/2} + c \eta \right) \: , \qquad \epsilon \geq \delta \: ,
\end{equation}
the decoder solves the $(n, a, m, \eta, \delta, \zeta, \epsilon , \mathcal{E})$-python's lunch decoding problem.
\end{theorem}

We defer the full proof until appendix \ref{app:ub}, but briefly outline the argument here. 
Set \(q=2^{-m}\). After adjoining the ancillas in the state \(|0^a\rangle\) and applying \(U\), the allowed inputs span the subspace
\begin{equation}
\mathcal S_{1}=U\bigl(\mathcal H_I\otimes|0^a\rangle_A\bigr) \: ,
\end{equation}
while successful post-selection corresponds to the subspace
\begin{equation}
\mathcal S_{2} =|0^m\rangle_P\otimes\mathcal H_O \: .
\end{equation}
The state-independent amplitude amplification iterate alternates two reflections. The first flips the sign of states in \(\mathcal S_{2}\). The second fixes \(\mathcal S_{1}\) and flips the sign of its orthogonal complement; it is implemented by conjugating the elementary reflection about \(\mathcal H_I\otimes|0^a\rangle_A\) by \(U\). As in Grover's algorithm, the product of these reflections rotates a state from \(\mathcal S_{1}\) toward its component in \(\mathcal S_{2}\). If \(K_U\) were exactly proportional to an isometry, every logical input would have the same overlap \(\sqrt q\) with \(\mathcal S_{2}\). The same \(O(q^{-1/2})=O(2^{m/2})\) iterations would therefore rotate every input close to its successfully postselected image, without the circuit needing to know the input state. Admissibility gives only approximate isometry \(q^{-1}K_U^\dagger K_U\approx\mathds1\), so that the worst-case decoding error cannot be made as small as in the perfectly isometric case. 
Reversing the resulting amplification circuit gives the decoder. The deviations from an exact isometry and from the optimal integer number of iterations produce the two terms lower bounding the worst-case decoding error in \eqref{eq:zeta_lb}. Since each iteration uses \(O(1)\) applications of \(U\) and \(U^\dagger\), together with polynomially complex reflections, the claimed circuit complexity bound follows.

It is natural to expect that the additional assumptions required for a lower bound $\mathcal{C}_{\text{PL}} = \Omega(2^{m/2})$ to hold should essentially be that there is no special structure in $U$ that can be exploited to reduce the complexity compared to the brute-force Grover method. 
In section \ref{sec:lower-bound}, we will prove a lower bound for quantum algorithms given query access to Haar random $U$, which provides some evidence for this idea.

\subsection{Admissible ensembles} \label{sec:admi}

To close this section, we will consider the role of the precise choice of admissible ensemble in the python's lunch conjecture. The main point we emphasize is that, while any admissible ensemble is suitable to establish an $O(2^{m/2})$ upper bound as per the result of the previous subsection, some admissible ensembles will not be consistent with a lower bound $\Omega(2^{m/2})$. 
This is entirely unsurprising, but reinforces the model-dependence of the conjecture. 
We will illustrate this point with the ensemble of unitaries highlighted by Brown et al. \cite{Brown:2019rox} in establishing the upper bound of the previous subsection, i.e. unitary 2-designs. 

First of all, we may establish that unitary 2-designs are admissible, and are therefore subject to the python's lunch complexity upper bound.\footnote{This admissibility result is weaker than the Haar admissibility theorem \ref{thm:Haar} as expected; using only the \(2\)-design property, our argument requires \(r=2^{n-m}\gg s^{2}/\eta^{2}\), whereas
the Haar case requires only \(r\gg s/\eta^{2}\).}

\begin{theorem}[Admissibility of unitary $2$-designs]
\label{thm:2design-admissible}
Let $\mathcal E$ be an exact unitary $2$-design on $n$ qubits, and
let
\begin{equation}
    d=2^n,\qquad
    s=2^{n-a},\qquad
    r=2^{n-m},\qquad
    q=\frac rd=2^{-m} \: .
\end{equation}
Then, for every $\eta>0$,
\begin{equation}
\label{eq:2design-admissible-prob}
    \Pr_{U\sim\mathcal E}\left[
        \left\|q^{-1}K_U^\dagger K_U-\mathds{1}_I\right\|
        >\eta
    \right]
    \leq
    \frac{s(d-r)(ds-1)}
    {\eta^2 r(d^2-1)}
    \leq
    \frac{s^2}{\eta^2r} \: .
\end{equation}
In particular, if $n - a + 2 \log(1/\eta) < a - m$, then $\mathcal E$ is $(a,m,\eta,\delta)$-admissible for
\begin{equation}
\label{eq:2design-delta}
    \delta
    =
    \frac{s^{2}}{\eta^2 r} \: .
\end{equation}
\end{theorem}

\begin{proof}
Set
\begin{equation}
    X_U \equiv K_U^\dagger K_U \: ,
    \qquad
    Q \equiv JJ^\dagger \: ,
\end{equation}
where $J = \mathds{1}_{I} \otimes |0^{a} \rangle_{A}$ denotes ancilla preparation, 
so that $Q$ is a rank $s$ projection. Since $\mathcal E$ is a
unitary $2$-design, we obtain
\begin{equation}
    \mathbb E_U [X_U] = q\,\mathds{1}_I \: .
\end{equation}
Similarly, for $A_U \equiv U^\dagger \Pi U$, the Haar second moment of a rank $r$
projection is
\begin{equation}
    \mathbb E_U[A_U\otimes A_U]
    =
    \frac{r(rd-1)}{d(d^2-1)}\,\mathds{1}
    +
    \frac{r(d-r)}{d(d^2-1)}\,F,
\end{equation}
where $F$ is the swap operator.  Since
\begin{equation}
    \Tr X_U^2
    =
    \Tr(A_UQA_UQ) \: ,
\end{equation}
it follows that
\begin{equation}
    \mathbb E_U [ \Tr X_U^2
    ] =
    \frac{sr\bigl[(rd-1)+(d-r)s\bigr]}
    {d(d^2-1)} \: .
\end{equation}
Consequently,
\begin{align}
    \mathbb E_U
    \left[\left\|q^{-1}X_U-\mathds{1}_I\right\|_F^2 \right]
    &=
    q^{-2}\mathbb E_U[\Tr X_U^2]-s \\
    &=
    \frac{s(d-r)(ds-1)}
    {r(d^2-1)} \: .
\end{align}
Since $\|\cdot\| \leq\|\cdot\|_F$, Markov's inequality gives
\begin{equation}
    \Pr_U\left[
    \left\|q^{-1}X_U-\mathds{1}_I\right\| >\eta
    \right] \leq
    \frac{1}{\eta^2} \:
    \mathbb E_U
    \left[\left\|q^{-1}X_U-\mathds{1}_I\right\|_F^2 \right] =
    \frac{s(d-r)(ds-1)}
    {\eta^2r(d^2-1)} \: .
\end{equation}
\end{proof}

It follows that, whenever $a - m \gg n - a$, the ensemble of holographic maps produced by taking $U$ to be a unitary 2-design allows for a decoder satisfying the python's lunch complexity upper bound.\footnote{Brown et al. \cite{Brown:2019rox} likewise construct a state-independent Grover decoder, but their \(2\)-design estimate is performed for an arbitrary fixed input state: averaging over $U$, they find an accumulated error that is small for $a\gg m\gg1$. This fixed-input estimate does not by itself imply that, with high probability over a single draw of \(U\), the decoder has small worst-case error simultaneously for all logical inputs. Our admissibility condition imposes precisely this condition, and our derivation introduces extra factors of the logical dimension $2^{n-a}$, leading to the more stringent sufficient condition $a-m\gg n-a$.}

On the other hand, the assumption that $\mathcal E$ is a unitary $2$-design (and is in particular admissible) is
not sufficient to guarantee $C_{\rm PL}=\Omega(2^{m/2})$. A convenient counter-example
is the uniform distribution over the $n$-qubit Clifford group, which is a
unitary $3$-design \cite{Webb2016,Zhu2017}.  We show in appendix \ref{app:Clifford}
that, with probability greater than $1-2^{m-a}$, the corresponding map $K_U$ is
(after rescaling) an exact stabilizer isometry and therefore admits a Clifford decoder of size
$O(n^2)$ due to a result of Cleve and Gottesman \cite{Cleve:1996pp}.

The efficient decoder does not rely on $U$ being a uniformly random $n$-qubit Clifford.  
It is enough that the final post-selected map $K_{U}$ be a stabilizer isometry, i.e., that it can (1) be implemented using Clifford gates, stabilizer ancillas, and stabilizer post-selections, and (2) that $K_{U}$ is an isometry. Isometric stabilizer tensor networks with post-selection provide a natural class of such holographic maps, though of course not every stabilizer tensor network with post-selection need be isometric.

\section{Query lower bound}
\label{sec:lower-bound}

We now turn to the query lower bound advertised in the previous section. 
In the following, it will be helpful to continue to denote $d = 2^{n}$, $q \equiv 2^{-m}$, and $r = dq = 2^{n-m}$. Our goal will be to establish the following theorem.

\begin{theorem}
\label{thm:query-lower-bound}
Suppose \(1\leq m<n\) and let $d = 2^{n}$.  
Let $U$ be Haar random in $U(d)$, and let $k$ be uniformly distributed in $\{0, 1\}$, independently of $U$. 
Let \(\mathcal A\) be a quantum algorithm that is
given one copy of \(\ket{\Psi_k(U)}\), 
makes at most \(T\) queries to \(U\) and \(U^\dagger\),
including controlled queries, and outputs a single bit. 
Then its average success probability (i.e. its probability of outputting $k$) satisfies
\begin{equation}
    \mathbb{E}[p_{\text{succ}}(U, \mathcal{A})] \leq
    \frac12 + 6\sqrt{2}\,T\,2^{-m/2}.
\end{equation}
Consequently, average success probability at least \(2/3\) requires
\begin{equation}
    T \geq \frac{1}{36\sqrt{2}}\,2^{m/2} \: ,
\end{equation}
and in particular, $T = \Omega(2^{m/2})$. 
\end{theorem}

A standard technique for proving quantum query lower bounds is the ``hybrid argument'', used in particular by Bennett, Bernstein, Brassard, and Vazirani [24] to establish the optimality of Grover's algorithm. In the form relevant here, the idea is that replacing an oracle \(U\) by another oracle \(U'\) in a \(T\)-query algorithm changes the final state by at most \(T\|U-U'\|\), provided the two computations begin in the same fiducial state. Consequently, oracles that are close in operator norm cannot be distinguished with high probability using few queries. There is a slight complication in applying this argument to our setting, which is that rather than extracting information purely from an oracle, we are attempting to extract information from both an oracle \(U\) and a supplied state \(|\Psi_k(U)\rangle\). 
The datum $k$ that we aim to extract is not encoded in $U$ at all, 
but rather in the correlation between $| \Psi_{k}(U) \rangle$ and $U$, which is a somewhat non-standard query complexity problem. 

To enable an application of the hybrid bound, our proof will proceed by constructing a joint distribution or ``coupling'' of two unitaries $U, U'$, each with a Haar random marginal, which are close together but satisfy\footnote{It is possible to simultaneously satisfy these conditions, i.e. closeness of $U$ to $U'$ and \eqref{eq:coupling_psis}, because $| \Psi_{k}(U) \rangle$ is defined via a post-selection, and represents an exponentially small component of $U | e_{k} (n) \rangle$. }
\begin{equation} \label{eq:coupling_psis}
    \ket{\Psi_0(U)} = \ket{\Psi_1(U')} \: .
\end{equation}
At this point, we can use a hybrid argument to show that an algorithm starting with $| \Psi_{0} (U) \rangle$ and querying $U$ would produce the same outcome with high probability as an algorithm starting with $| \Psi_{1} (U') \rangle$ and querying $U'$, unless the number of queries is sufficiently large. Since $U, U'$ are both marginally Haar random, a successful algorithm should output zero in the first case and one in the second case with high probability, giving a contradiction unless the number of queries is large. 

\subsection{Coupling of Haar random two-frames}

Our eventual goal will be to construct a coupling of Haar random unitaries $(U, U')$ with the property $\Pi U | e_{0}(n) \rangle = \Pi U' |e_{1}(n) \rangle$.  
To achieve this, we will begin by constructing a coupling of random \emph{two-frames} $(C, C')$,\footnote{A $k$-frame is simply a $k$-tuple of linearly independent vectors in a vector space. Here, we consider orthonormal two-frames in the $d$-dimensional Hilbert space $\mathbb{C}^{d}$. Of course, an orthonormal $d$-frame in this Hilbert space defines a unitary, by the identification of the $k^{\textnormal{th}}$ element of the frame with $U | k \rangle$. } then complete these to unitaries. 

\subsubsection*{Structure of random two-frame \texorpdfstring{$C$}{}}

To this end, define the two-frame
\begin{equation}
    C \equiv ( U |e_0 (n) \rangle \: , \: U |e_1 (n)\rangle ) \: , \qquad C^{\dagger} C = \mathds{1}_{2} \: .
\end{equation}
This two-frame is distributed according to the uniform measure on the ``Stiefel manifold'' $V_{2}(\mathbb{C}^{d})$ (i.e. the manifold of orthonormal two-frames in $\mathbb{C}^{d}$). 
Our goal in the next few paragraphs will be to understand the statistical properties of $C$ with respect to the post-selection subspace and its orthogonal complement; this will be crucial for defining a second two-frame $C'$ with the desired properties.

We may decompose $C$ according to the decomposition of the Hilbert space $\mathbb{C}^{d}$ into the support of the post-selection projector $\Pi$ and its orthogonal complement, 
\begin{equation}
    \mathbb{C}^d
    =
    \Pi \mathbb{C}^d
    \oplus
    (\mathds{1}_{d}-\Pi )\mathbb{C}^d \: ,
\end{equation}
defining
\begin{equation}
    C_{\parallel} \equiv \Pi C \: , 
    \qquad
    C_{\perp} \equiv (\mathds{1}_{d} - \Pi) C \: , 
    \qquad
    G \equiv C_{\parallel}^\dagger C_{\parallel} = C^\dagger \Pi C \: .
\end{equation}
Since \(C^\dagger C=\mathds{1}_2\), it follows that
\begin{equation}
    C_{\perp}^\dagger C_{\perp} = \mathds{1}_2 - G \: .
\end{equation}
The objects $C_{\parallel}$ and $C_{\perp}$ are just the components of $C$ in the post-selected subspace and its orthogonal complement. 

Now, because $\text{rk}(\Pi) = r$ and $\text{rk}(\mathds{1}_{d} - \Pi) = d-r$ are at least two, the matrices $C_{\parallel}, C_{\perp}$ will almost surely have trivial kernel, and both $G$ and $\mathds{1}_{2} - G$ will almost surely be positive definite. 
We may generally write the polar
decompositions
\begin{equation}
    C_{\parallel} = V \sqrt{G} \: ,
    \qquad
    C_{\perp} = W \sqrt{\mathds{1}_2-G} \: ,
\end{equation}
where $V, W$ are partial isometries;\footnote{That is, they are isometries on the complement of their kernels.} when $G, \mathds{1}_{2} - G$ are positive definite, they are in fact isometries
\begin{equation}
    V^\dagger V = W^\dagger W = \mathds{1}_2 \: ,
\end{equation}
given explicitly by 
\begin{equation}
    V = C_{\parallel} G^{-1/2} \: , \qquad W = C_{\perp} (\mathds{1}_{2} - G)^{-1/2} \: .
\end{equation}
From this, it is clear that the range of \(V\) lies in \(\Pi \mathbb{C}^d\), and the range of \(W\) lies
in \((\mathds{1}_{d}-\Pi)\mathbb{C}^d\), as expected for the polar constituents of $C_{\parallel}, C_{\perp}$.\footnote{This statement holds when $V, W$ are exact isometries; when they are merely partial isometries, it continues to hold provided we take $V$ and $W$ to be the canonical partial isometries which vanish on the kernels of $C_{\parallel}$ and $C_{\perp}$ respectively. } We thus have
\begin{equation}
    C = V \sqrt{G} + W \sqrt{\mathds{1}_2 - G} \: ,
\end{equation}
with $(V, W) \in V_{2}(\mathbb{C}^{r}) \times V_{2}(\mathbb{C}^{d-r})$. 

We will be interested in understanding the distributions of the constituents $G, V, W$; our main conclusion will be that the distribution of $(V, W)$ is in fact simply the product of uniform distributions, and is independent of $G$.  
First of all, for every
\(R\in U(2)\), right-invariance of the uniform measure on two-frames implies
\begin{equation}
    C\overset{d}{=}CR \: ,
\end{equation}
where the equality with superscript $d$ denotes that the left and right hand side are identically distributed, and therefore
\begin{equation}
    G\overset{d}{=}R^\dagger GR \: .
\end{equation}
Thus, the distribution of \(G\) is invariant under unitary conjugation.

Second of all, 
we observe that for arbitrary
\begin{equation}
    L_{\parallel} \in U \left( \Pi \mathbb{C}^d \right),
    \qquad
    L_{\perp} \in U\left( (\mathds{1}_{d}-\Pi)\mathbb{C}^d \right) \: ,
\end{equation}
i.e. unitaries on the indicated subspaces, the transformation
\begin{equation}
    C \rightarrow (L_{\parallel} \oplus L_{\perp}) C
\end{equation}
leaves \(G\) invariant, 
and sends
\begin{equation} \label{eq:VWtrans}
    (V,W) \rightarrow (L_{\parallel} V \: , \: L_{\perp}W) \: .
\end{equation}
One has
\begin{equation}
    C \overset{d}{=} (L_{\parallel} \oplus L_{\perp}) C
\end{equation}
due to left-invariance of the uniform measure, from which it follows that
\begin{equation}
    \begin{split}
        V & = \Pi C G^{-1/2} \overset{d}{=} \Pi (L_{\parallel} \oplus L_{\perp}) C G^{-1/2} = L_{\parallel} V \\
        W & = (\mathds{1}_{d} - \Pi) C (\mathds{1}_{2} - G)^{-1/2} \overset{d}{=} (\mathds{1}_{d} - \Pi) (L_{\parallel} \oplus L_{\perp}) C (\mathds{1}_{2} - G)^{-1/2} = L_{\perp} W \: ,
    \end{split}
\end{equation}
that is,
\begin{equation}
    (G, V, W) \overset{d}{=} (G, L_{\parallel} V, L_{\perp} W) \: .
\end{equation}
We conclude that the distribution of $(V, W)$ conditioned on fixed $G$ is invariant under \eqref{eq:VWtrans}, and that it must therefore be the product of uniform distributions on $V_{2}(\mathbb{C}^{r})$ and $V_{2}(\mathbb{C}^{d-r})$. 
In particular, the distribution of $(V, W)$ is in fact independent of $G$.

\subsubsection*{Construction of two-frame \texorpdfstring{$C'$}{}}

We can now construct the two-frame $C'$. The basic idea is to leave the isometry $W$ associated with the orthogonal complement $(\mathds{1}_d-\Pi)\mathbb{C}^d$ unchanged, while exchanging the two columns of the isometry $V$ associated with the post-selected subspace $\Pi\mathbb{C}^d$. Since a Haar-random two-frame has only a small overlap with the post-selected subspace when $m$ is large, this will allow us to keep $C$ and $C'$ close together while modifying their post-selected components. In fact, we will also need to transform the Gram matrix $G$ to ensure that the full post-selected components are exchanged exactly. As we will see, this can be done while keeping $C$ and $C'$ close on average, and preserving their uniform marginal distributions.

Now, denote
\begin{equation}
    X \equiv 
    \begin{pmatrix}
        0 & 1 \\
        1 & 0
    \end{pmatrix} \: ,
\end{equation}
and define
\begin{equation}
    G' \equiv XGX \: ,
    \qquad
    V' \equiv VX \: ,
    \qquad
    W' \equiv W \: ,
\end{equation}
and
\begin{equation}
    C'
    \equiv 
    V' \sqrt{G'}
    +
    W' \sqrt{\mathds{1}_2-G'} \: .
\end{equation}
Since functional calculus commutes with unitary conjugation,
\begin{equation}
    \sqrt{G'} = X \sqrt{G} X \: ,
    \qquad
    \sqrt{\mathds{1}_2-G'} = X \sqrt{\mathds{1}_2-G} X \: ,
\end{equation}
and
\begin{equation}
    C'
    =
    V \sqrt{G} X
    +
    W X \sqrt{\mathds{1}_2-G} X \: .
\end{equation}

The joint distribution of \((G',V',W') = (XGX, VX, W)\) is the same as that of \((G,V,W)\):
indeed, the distribution of $(VX, W)$ must be independent of $XGX$ if $(V, W)$ is independent of $G$, and we know that $G \overset{d}{=} XGX$ and $(V, W) \overset{d}{=} (VX, W)$ by invariance properties of uniform measures. 
Consequently,
\begin{equation}
    C'\overset{d}{=}C \: ,
\end{equation}
and \(C'\) is also a uniformly random orthonormal two-frame. 
Moreover,
\begin{equation}
    \Pi C'
    =
    V \sqrt{G} X
    =
    (\Pi C)X \: ,
\end{equation}
so in particular
\begin{equation}
    \Pi C' |e_1 (1) \rangle = \Pi C | e_0 (1) \rangle \: , \qquad \Pi C' | e_0 (1) \rangle = \Pi C | e_1 (1) \rangle \: .
\end{equation}
Thus, the normalized post-selected state associated with the second column
of \(C'\) is exactly the normalized post-selected state associated with
the first column of \(C\), and vice versa. 

\subsubsection*{Distance bound}

Despite the fact that $\Pi C, \Pi C'$ are related by interchanging their columns, we can nevertheless show that $C, C'$ are close in operator norm on average. Intuitively, $C, C'$ are only swapping a small $O(\sqrt{q})$ component in the direction of the post-selected subspace, and remain close on the orthogonal subspace. We will proceed to bound the distance between \(C\) and \(C'\).  Let
\begin{equation}
    H \equiv \sqrt{\mathds{1}_2-G} ,
    \qquad
    g \equiv \lVert G \rVert \: .
\end{equation}
Then
\begin{equation}
    C-C'
    =
    V \sqrt{G} (\mathds{1}_2-X)
    +
    W (H-XHX) \: .
\end{equation}
Since \(\| \mathds{1}_2-X\| = 2\), sub-multiplicativity gives
\begin{equation}
    \|V \sqrt{ G} (\mathds{1}_2-X)\|
    \leq 2\sqrt{g} \: .
\end{equation}
Also, by triangle and invariance of the norm under unitary conjugation, 
\begin{equation}
    \|H-XHX\|
    \leq
    \|H-\mathds{1}_2\| +\|\mathds{1}_2-XHX\| =2\|H- \mathds{1}_2\| \: .
\end{equation}
One has
\begin{equation}
    \|H-\mathds{1}_2\|
    =
    1-\sqrt{1-g}
    \leq \sqrt{g} \: ,
\end{equation}
given that $g \in [0, 1]$, so all together, 
\begin{equation}
    \|C-C'\| \leq 4\sqrt{g}.
\end{equation}

By taking an expectation value with respect to the distribution of $G$, we can more directly relate this to Hilbert space dimensions. 
Since \(G\geq 0\),
\begin{equation}
    g\leq \text{tr} (G) \: .
\end{equation}
Moreover,
\begin{align*}
    \mathbb{E}[ \text{tr} (G)]
    &= \mathbb{E}[ \langle e_{0}(1)| C^{\dagger} \Pi C |e_{0} (1) \rangle + \langle e_{1} (1) | C^{\dagger} \Pi C |e_{1} (1) \rangle ]
    \\
    & = \mathbb{E} \left[ 
        \| \Pi U | e_0 (n) \rangle \|^2+\| \Pi U |e_1 (n) \rangle \|^2 \right] = 2q \: .
\end{align*}
By Jensen's inequality,
\begin{equation}
    \mathbb{E}[\sqrt{g}]
    \leq
    \sqrt{\mathbb{E}[g]}
    \leq
    \sqrt{2q} \: ,
\end{equation}
hence
\begin{equation}
\label{eq:frame-distance}
    \mathbb{E} \left[ \|C-C'\| \right]
    \leq
    4\sqrt{2q} \: .
\end{equation}
We have therefore managed to construct coupled two-frames $C, C'$ which are close in operator norm on average, with the property that the projection of the first vector in $C$ and the second vector in $C'$ onto the subspace $\Pi \mathbb{C}^{d}$ agree. 

\subsection{Completing the two-frames to Haar random unitaries}

To proceed, we use the following lemma, whose proof is provided in appendix \ref{app:lemmas}. 

\begin{lemma}
\label{lem:frame-extension}
Let \(C,C':\mathbb{C}^2\to\mathbb{C}^d\) be isometries.  Then there exists a
unitary \(L\in U(d)\) such that
\begin{equation}
    LC=C'
\end{equation}
and
\begin{equation}
    \|L- \mathds{1}_{d} \| \leq 3\|C-C'\| \: .
\end{equation}
\end{lemma}

We now complete \(C\) to a Haar random unitary \(U\).  Conditional on
\(C\), choose a uniformly random orthonormal completion \(D\), and write
\begin{equation}
    U=\begin{bmatrix}C&D\end{bmatrix} \: .
\end{equation}
Let \(L\) be as in lemma~\ref{lem:frame-extension}, and define
\begin{equation}
    U' \equiv LU \: .
\end{equation}
The first two columns of \(U'\) are exactly \(C'\), which are uniformly distributed on $V_{2}(\mathbb{C}^{d})$. Moreover, conditioned on those columns, the remaining columns are uniformly distributed on $V_{2}(\mathbb{C}^{d})$. It follows that $U'$ is itself Haar random,\footnote{It is not sufficient to infer that $U'$ is Haar random from $U' = LU$, since $L$ depends on $C, C'$. } i.e. $U, U'$ form a coupling of Haar random unitaries. 
Furthermore,
\begin{equation}
    \Pi U' | e_1 (n) \rangle = \Pi U | e_0 (n) \rangle \: ,
\end{equation}
and hence
\begin{equation}
\label{eq:identical-inputs}
    \ket{\Psi_1(U')}=\ket{\Psi_0(U)} \: .
\end{equation}
Finally,
\begin{equation}
    \|U-U'\|
    =
    \|(\mathds{1}_{d} -L)U\|
    =
    \|\mathds{1}_{d}-L\|
    \leq
    3\|C-C'\| \: .
\end{equation}
Using \eqref{eq:frame-distance}, this implies
\begin{equation}
\label{eq:oracle-distance}
    \mathbb{E}\left[ \|U-U'\| \right] 
    \leq
    12\sqrt{2q} \: .
\end{equation}

\subsection{Hybrid argument}

We are now ready to apply a hybrid argument to upper bound the success probability of any distinguishing algorithm querying $U, U^{\dagger}$. 
Let the probability of outputting zero given input $i$ and unitary $U$ be
\begin{equation}
    p_{0}^{(i)} (U, \mathcal{A})
    \equiv
    \Pr\!\left[
        \mathcal A\text{ outputs }0
        \,\middle|\,
        (U \: , \: \ket{\Psi_{i}(U)})
    \right] \: , \qquad i \in \{0, 1\} \: .
\end{equation}
The average success probability is
\begin{equation}
    p_{\mathrm{succ}}(\mathcal{A})
    =
    \frac12\mathbb{E}[p_{0}^{(0)} (U, \mathcal{A})
    ]
    +
    \frac12\mathbb{E}[ 1-p_{0}^{(1)} (U, \mathcal{A})
    ] \: ,
\end{equation}
and therefore
\begin{equation}
\label{eq:advantage}
    2\left(p_{\mathrm{succ}}(\mathcal{A}) - \frac12\right)
    =
    \mathbb{E}[p_{0}^{(0)} (U, \mathcal{A})
    ] -\mathbb{E}[p_{0}^{(1)} (U, \mathcal{A})
    ] \: .
\end{equation}

We use the standard oracle hybrid bound. 
Suppose two executions of a
\(T\)-query algorithm begin in the same state and use the oracles \(U\)
and \(U'\), respectively.  After purifying the algorithm and deferring
all measurements, the usual telescoping argument gives that the (pure) state $| \phi_{T}(U) \rangle$ after $T$ queries to $U$ or $U^{\dagger}$ satisfies
\begin{equation}
    \bigl\|
        \ket{\phi_T(U)}-\ket{\phi_T(U')}
    \bigr\|_{2}
    \leq
    T\|U-U'\| \: . 
\end{equation}
The same estimate holds if individual queries are to \(U^\dagger\), 
and also for controlled queries.  The difference in the probability of
any final measurement outcome is therefore at most $T\|U-U'\|$; in particular, this upper bounds the difference in the probabilities of the algorithm outputting zero on some input state when querying $U$ or $U'$.  

We now apply this bound to the coupling of Haar random unitaries constructed above. By
\eqref{eq:identical-inputs}, the two executions corresponding to
\begin{equation}
    \left( \ket{\Psi_0(U)},U \right)
    \qquad\text{and}\qquad
    \left( \ket{\Psi_1(U')},U' \right)
\end{equation}
begin with exactly the same supplied state.  Hence
\begin{equation}
    | p_{0}^{(0)} (U, \mathcal{A})
    - p_{0}^{(1)} (U', \mathcal{A})
     |
    \leq
    T\|U-U'\| \: .
\end{equation}
Moreover, because \(U'\) is marginally Haar-random,
\begin{equation}
    \mathbb{E}[p_{0}^{(1)} (U', \mathcal{A})
    ]=\mathbb{E}[p_{0}^{(1)} (U, \mathcal{A})
    ].
\end{equation}
Combining this with \eqref{eq:oracle-distance} and
\eqref{eq:advantage}, we obtain
\begin{equation}
    \begin{split}
    2\left(p_{\mathrm{succ}}(\mathcal{A})-\frac12\right)
    &=
    \mathbb{E} [ p_{0}^{(0)} (U, \mathcal{A})
    -p_{0}^{(1)} (U', \mathcal{A})
     ] \leq
    \mathbb{E}\left[ |p_{0}^{(0)} (U, \mathcal{A})
    -p_{0}^{(1)} (U', \mathcal{A})| \right] \\
    &\leq
    T\,\mathbb{E}[ \|U-U'\| ] \leq
    12\sqrt{2}\,T\,\sqrt q \: .
    \end{split}
\end{equation}
Recalling that \(q=2^{-m}\),
\begin{equation}
    p_{\mathrm{succ}}(\mathcal{A})
    \leq
    \frac12
    +
    6\sqrt{2}\,T\,2^{-m/2}.
\end{equation}
This proves theorem~\ref{thm:query-lower-bound}.
In particular, it is immediate that if \(p_{\mathrm{succ}}(\mathcal{A}) \geq 2/3\), then
\begin{equation}
    T
    \geq
    \frac{1}{36\sqrt{2}}\,2^{m/2} \: .
\end{equation}

\qed

\subsection{Lower bounds in other admissible ensembles}

The query lower bound of the last section only directly applies to the case that the unitary circuit ensemble $\mathcal{E}$ is Haar random, or at least indistinguishable from Haar even using a number of queries which is exponentially large in $m$. Extending it directly to other ensembles would require controlling the distinguishability of these ensembles from Haar randomness for algorithms making exponentially many oracle queries, with the additional complication that our decoder also receives a state dependent on the oracle. This is a very strong requirement, and is unlikely to hold for physically motivated families of unitaries based on holographic tensor networks.

To emphasize this point, consider an ensemble of unitary quantum circuits built from all-to-all two-qubit gates, with \(O(n)\) gates per layer. Brandão, Harrow, and Horodecki \cite{Brandao:2012zoj} established a lower bound on the circuit size required to generate sufficiently accurate unitary \(T\)-designs; more recently, Schuster, Ma, Lombardi, Brandão, and Huang \cite{Schuster:2025vta} stated a corresponding lower bound for operationally defined strong unitary \(T\)-designs, which remain indistinguishable from Haar under arbitrary \(T\)-query experiments. For circuits with at most \(O(n)\) ancilla qubits, this gives the depth bound
\begin{equation}
\ell=\Omega\left(\frac{T}{\log(nT)}\right) \: .
\end{equation}
In particular, when \(T=\Theta(2^{m/2})\) and \(m=\Theta(n)\), the required depth is exponential in \(n\). This is consequently not a particularly physically well-motivated assumption for holographic states, except perhaps for the exponentially late-time behaviour of such states. 

For structured polynomial depth tensor networks besides brickwork circuits, recent results on learning unknown unitaries actually provide upper bounds on query complexity. For example, consider the one-bit recovery task for a python's lunch tensor network with $N$ independent Haar random tensors $U\in U(d)$ at each node of some generic fixed structure. Suppose we allow query access to each individual Haar random tensor, and the network structure is known. Adapting the results of \cite{haah2023query}, we then find an upper bound of $$\tilde{O}_{\beta,\varepsilon}(N^3d^{7/2}\log N)$$ queries suffices to recover the one bit (with some error allowance notated by $\varepsilon$ and $\beta$). The protocol proceeds by learning classical descriptions of each of the individual tensors, classically determining a measurement that would optimally distinguish the two possibilities using the known structure of the tensor network, and then implementing that measurement on the given unknown state. For completeness, we describe a detailed analysis of this process in appendix \ref{app:learningbound}.

Notably this query upper bound has no dependence on the number of post-selected qubits $m$. This occurs because the entirety of the query cost comes from learning each individual unitary; the classical operations performed in subsequent steps to analyze the effects of post-selection do not factor into the query complexity, though they of course contribute significantly to the gate complexity. As such, this example illustrates the obstructions that query complexity techniques seem likely to run into for the setting of structured circuits.

\section{Discussion}
\label{sec:discussion}

We have established a tight \emph{query-complexity} characterization of the python's lunch decoding problem when the unitary \(U\) is Haar random. The amplitude amplification construction of Brown et al. \cite{Brown:2019rox} gives an \(O(2^{m/2})\)-query decoder for any approximately isometric holographic map of the form $K_{U}$, while our argument shows that a map $K_{U}$ built from Haar random $U$ requires \(\Omega(2^{m/2})\) queries even to recover a single logical bit with constant advantage. Thus, in the black-box setting, the $\Omega(2^{m/2})$ scaling is optimal.
We emphasize, of course, that this does not preclude more efficient decoding algorithms that use additional structure of the encoder, or applicable to different ensembles of holographic tensor networks.

\subsubsection*{The python's lunch conjecture and ensembles of holographic maps}

Our results 
have also emphasized that the same amount of post-selection \(m\) can occur in ensembles of holographic maps with very different decoding complexities. Whereas Haar random unitaries exhibit the optimal black box lower bound derived here, other random ensembles, e.g. random Cliffords, may admit vastly more efficient decoders. 
This delineation between different ensembles is unsurprising, but it suggests some caution may be warranted in interpreting the python's lunch formula as a statement determined by geometry alone. 

A related point has been made recently by Arora et al.~\cite{Arora:2026crs}. They show that the quantitative python's lunch prediction is not automatic in generic random tensor network models: their exponential complexity need not be controlled by the usual python's lunch exponent. They attribute this failure to the absence of \emph{computational covariance}, meaning, roughly, that the network cannot be decomposed along arbitrary suitable bulk slices into low-complexity maps. They then construct twirled perfect tensor networks possessing this property, and recover an upper bound governed by the python's lunch value.

That work and the present result address complementary aspects of the conjecture. Computational covariance concerns the structure needed to isolate post-selection as the expensive part of a holographic tensor network. Our work 
does not assume such a structure, but instead considers maps with the ``unitary followed by post-selection'' form of \eqref{eq:intro_KU} and asks whether the resulting \(2^{m/2}\) black box cost can be improved. For Haar random \(U\), it cannot. At the same time, the Haar ensemble lacks the efficient local description expected of a physical holographic map (let alone a computationally covariant one), and our result does not settle the corresponding circuit complexity question.

A more complete formulation of the python's lunch conjecture therefore appears to require assumptions of two different kinds. The holographic maps should possess enough geometric or computational covariance to admit a decomposition into efficiently implementable pieces separated by post-selection. They must also be sufficiently computationally generic that the
decoder cannot be synthesized by exploiting special structure in those pieces. Stabilizer tensor networks satisfy the first requirement but fail the second, while Haar random maps provide the strongest form of genericity but do not have an efficient microscopic description.
An important open problem is to establish super-polynomial query complexity lower bounds for physically motivated ensembles between these extremes. 

\subsubsection*{Acknowledgments}

We thank Adam Bouland for posing the question of a query lower bound to the author, and Tamara Kohler and Alex May for discussions at the onset of this project. 
We also thank Alex May for helpful comments on the draft. Research at the Perimeter Institute is supported by the Government of Canada through the Department of Innovation, Science and Industry Canada and by the Province of Ontario
through the Ministry of Colleges and Universities.
Many of the proofs in this work, including some ideas in the proof of theorem \ref{thm:query-lower-bound}, were assisted by ChatGPT 5.6.

\appendix

\section{Admissibility of the Haar ensemble} \label{app:admissibility}

In this appendix, we provide a proof of the admissibility of the Haar ensemble, as stated in theorem \ref{thm:Haar} of section \ref{sec:pld-problem}. 

\begin{mythm}{1}
    The Haar ensemble on unitaries $U(2^{n})$ is $(a, m, \eta, \delta)$-admissible whenever $0 < \eta \leq 1$ and
    \begin{equation}
        2^{n - m} \geq C \eta^{-2} \left( 2^{n-a} + c_{1} \ln(1 / \delta) + c_{2} \right) 
    \end{equation}
    for some universal constants $C, c_{1}, c_{2} >0$. 
\end{mythm}

\begin{proof}
    Recall that an \emph{$\epsilon$-net} for a Hilbert space $\mathcal{H}$ is a set $S$ such that, for every state $| \psi \rangle \in \mathcal{H}$, there exists $| \phi \rangle \in S$ such that
    \begin{equation}
        \| | \psi \rangle - | \phi \rangle \|_{2} \leq \epsilon \: ;
    \end{equation}
    moreover, an $\epsilon$-net is called \emph{minimal} if removing any element gives a set $S'$ which is no longer an $\epsilon$-net. 
    This definition is provided in e.g. appendix D of \cite{Akers:2022qdl}, which also notes that any $\epsilon$-net $S$ must satisfy
    \begin{equation}
        | S| \geq \sqrt{\pi d} \epsilon^{-2d} \: ,
    \end{equation}
    and there always exists a minimal $\epsilon$-net $S_{*}$ such that
    \begin{equation}
        | S_{*} | \leq \left( 1 + \frac{2}{\epsilon} \right)^{2d} \: .
    \end{equation} 
    We may therefore choose a $\frac{1}{4}$-net $S$ for our $(2^{n-a})$-dimensional Hilbert space $\mathcal{H}$ which satisfies $|S| \leq 9^{2^{n-a+1}}$. 

    For any $| \psi \rangle \in \mathcal{H}$ and $0 < \epsilon \leq 1$, one can show that
    \begin{equation} \label{eq:fact2}
        \underset{U \sim \text{Haar}}{\text{Pr}} \left[ \| K_{U}| \psi \rangle \|_{2} \notin [(1 - \epsilon) 2^{-m/2}, (1 + \epsilon) 2^{-m/2}] \right] \leq 4 \exp \left( - \frac{ 2^{n-m}\epsilon^{2}}{16} \right) \: . 
    \end{equation}
    This result is closely related to the Johnson-Lindenstrauss lemma, and is stated e.g. as ``Fact 2'' in \cite{Sen:2018ivg}; see e.g. \cite{Meckes2019RandomMatrix} for related discussion. 

    Now, taking $\epsilon = \eta / 6$, and denoting $H_{U} = 2^{m} K_{U}^{\dagger} K_{U} - \mathds{1}$ for convenience, we have that the complementary event to the one whose probability is evaluated in \eqref{eq:fact2} satisfies
    \begin{equation}
        | \langle \psi | H_{U} | \psi \rangle | = | 2^{m} \| K_{U} | \psi \rangle \|_{2} - 1 | \leq 2 \epsilon + \epsilon^{2} < \frac{\eta}{2} \: . 
    \end{equation}
    Consequently, a union bound over $| \psi \rangle \in S$ gives
    \begin{equation}
        \begin{split}
            \underset{U \sim \text{Haar}}{\text{Pr}} \left[ \max_{| \psi \rangle \in S} | \langle \psi | H_{U} | \psi \rangle | \geq \frac{\eta}{2} \right] & \leq 4 |S| \exp \left( - \frac{\eta^{2} 2^{n-m}}{576} \right) \\
            & \leq 4 \times 9^{2^{n-a+1}} \times \exp \left( - \frac{\eta^{2} 2^{n-m}}{576} \right) \: .
        \end{split}
    \end{equation}

    Finally, we note that for any Hermitian $H$,
    \begin{equation}
        \| H \| \leq 2 \max_{| \psi \rangle \in S} | \langle \psi | H | \psi \rangle | \: .
    \end{equation}
    This is because one can choose unit eigenvector $| \phi \rangle$ for $H$ such that $\| H \| = | \langle \psi | H | \psi \rangle|$, and there exists $| \psi \rangle \in S$ such that $\| | \psi \rangle - | \phi \rangle \|_{2} \leq \frac{1}{4}$ by the $\frac{1}{4}$-net property, so
    \begin{equation}
        \begin{split}
            | \langle \psi | H | \psi \rangle - \langle \phi | H | \phi \rangle | & \leq \| | \psi \rangle - | \phi \rangle \|_{2} \cdot \| H \| \cdot \| | \psi \rangle \|_{2} + \| | \phi \rangle \|_{2} \cdot  \| H \| \cdot \| | \psi \rangle - | \phi \rangle \|_{2} \\
            & \leq \frac{1}{2} \| H \| \: ,
        \end{split}
    \end{equation}
    and thus
    \begin{equation}
        | \langle \phi | H | \phi \rangle | \geq | \langle \psi | H | \psi \rangle | - | \langle \psi | H | \psi \rangle - \langle \phi | H | \phi \rangle | \geq \frac{1}{2} \| H \| \: .
    \end{equation}
    We therefore have 
    \begin{equation}
        \underset{U \sim \text{Haar}}{\text{Pr}} \left[ \| H_{U} \| \leq \eta \right] \geq 1 -  4 \times 9^{2^{n-a+1}} \times \exp \left( - \frac{\eta^{2} 2^{n-m}}{576} \right) \: ,
    \end{equation}
    meaning that the Haar ensemble is admissible provided
    \begin{equation}
        \delta > 4 \times 9^{2^{n-a+1}} \times \exp \left( - \frac{\eta^{2} 2^{n-m}}{576} \right) \: .
    \end{equation}
    Taking a logarithm of both sides and rearranging,
    \begin{equation}
        2^{n-m} > (1152 \times \ln(9)) \times  \eta^{-2} \left( 2^{n-a} + \frac{\ln(4/ \delta)}{2 \ln(9)} \right) \: ,
    \end{equation}
    as desired. 
\end{proof}

\section{Proof of complexity upper bound}
\label{app:ub}

We will provide a proof of the complexity upper bound in theorem \ref{thm:grover-upper-bound} of section \ref{sec:ub} due to the state-independent amplitude amplification procedure of \cite{Brown:2019rox}. 

\begin{proof}
Let
\begin{equation}
    q \equiv 2^{-m} \: , \qquad
    J \equiv \mathds{1}_I \otimes \ket{0^a}_A \: ,
    \qquad
    \Pi \equiv \ket{0^m}\!\bra{0^m}_P\otimes\mathds{1}_O \: ,
\end{equation}
so that
\begin{equation}
    K_U =(\bra{0^m}_P\otimes\mathds{1}_O)UJ \: ,
    \qquad
    \widetilde K_U \equiv \Pi UJ
       =\ket{0^m}_P\otimes K_U \: .
\end{equation}
Recall that, on the admissibility event,
\begin{equation}
    \left\| q^{-1} K_U^\dagger K_U-\mathds{1}_I\right\|
    \leq\eta \: .
\end{equation}
Writing the polar decomposition
\begin{equation}
    K_U=V_US_U,
    \qquad
    S_U=(K_U^\dagger K_U)^{1/2} \: ,
\end{equation}
and normalizing the positive part $T_U=q^{-1/2}S_U$, we therefore have\footnote{We use that admissibility gives $\| T_{U}^{2} - \mathds{1}_{I} \| \leq \eta$, which together with positivity of $T_{U}$ implies $\sqrt{1-\eta} \leq t \leq \sqrt{1 + \eta}$ for every eigenvalue $t$ of $T_{U}$. Provided $0 < \eta < 1$, it follows that $\| T_{U} - \mathds{1}_{I} \| \leq 1 - \sqrt{ 1 - \eta} = \Delta_{\eta}$. }
\begin{equation}
    \|T_U-\mathds{1}_I\| \leq \frac{\eta}{1 + \sqrt{1 - \eta}} \equiv \Delta_{\eta} \: .
\end{equation}
Hence, for all states $\ket{\psi}_{IR}$, one has\footnote{Define $| \tilde{\chi} \rangle \equiv (T_{U} \otimes \mathds{1}_{R}) | \psi \rangle$ and $| \chi \rangle = | \tilde{\chi} \rangle / \| | \tilde{\chi} \rangle \|$. Then 
\begin{equation}
    | \Psi(\psi, U) \rangle = \frac{(V_{U} \otimes\mathds{1}_R)\ket{\tilde{\chi}}}
    {\|(V_{U} \otimes\mathds{1}_R)\ket{\tilde{\chi}}\|_{2}} = (V_{U} \otimes\mathds{1}_R)\ket{\chi} \: ,
\end{equation}
and thus
\begin{equation}
    \Bigg\| | \Psi(\psi, U) \rangle  - (V_{U} \otimes \mathds{1}_{R}) | \psi \rangle \Bigg\|_{2} = \| | \psi \rangle - | \chi \rangle \|_{2} \leq \| | \psi \rangle - | \tilde{\chi} \rangle \|_{2} + \| | \tilde{\chi} \rangle - | \chi \rangle \|_{2} \: .
\end{equation} 
The second term on the right hand satisfies
\begin{equation}
    \| | \tilde{\chi} \rangle - | \chi \rangle \|_{2} = | 1 - \| | \tilde{\chi} \rangle \|_{2} | \leq \| | \psi \rangle - | \tilde{\chi} \rangle \|_{2}
\end{equation}
by reverse triangle, and $\| | \psi \rangle - | \tilde{\chi} \rangle \|_{2}$ is upper bounded by $\| T_{U} - \mathds{1}_{I}\|$.  
}
\begin{equation}
\label{eq:normalized-close-to-polar}
    \| | \Psi(\psi, U) \rangle
    -
    (V_U\otimes\mathds{1}_R)\ket{\psi} \|_{2}
    \leq 2 \Delta_{\eta} \: .
\end{equation}
where we recall 
that $| \Psi(\psi, U) \rangle$ is the normalized post-selected state.

Consequently, if we can construct a unitary circuit $Q_{U}$ from $IA$ to $OP$ satisfying
\begin{equation}
    \| Q_{U} J - |0^{m} \rangle_{P} \otimes V_{U} \| \leq \alpha \: ,
\end{equation}
then $\mathcal{D}_{U} \equiv \text{tr}_{A} \circ Q_{U}^{\dagger}$ will be a good decoder in the sense that
\begin{equation}
    \text{err}(U, \mathcal{D}_{U}) \leq \| (Q_{U}^{\dagger} \otimes \mathds{1}_{R}) (|0^{m} \rangle_{P} \otimes | \Psi(\psi, U) \rangle) - J | \psi \rangle \|_{2} \leq 2 \Delta_{\eta} + \alpha \: .
\end{equation}
This can be done with
\begin{equation}
    \alpha = \alpha_{q, \eta} \equiv \sin^{-1}\sqrt{q} + \frac{\pi/2 + \sqrt{q}}{\sqrt{1 - q(1 + \eta)}} \times \Delta_{\eta}
\end{equation}
by the state-independent Grover protocol of \cite{Brown:2019rox}.

Indeed, let
\begin{equation}
    R_J \equiv 2JJ^\dagger-\mathds{1} \: ,
    \qquad
    R_{\Pi} \equiv \mathds{1}-2 \Pi \: ,
\end{equation}
which are evidently unitary given that $J J^{\dagger}$ and $\Pi$ are projections, 
and consider the Grover iterate
\begin{equation}
    M_U \equiv UR_JU^\dagger R_\Pi \: .
\end{equation}
Let $\sigma_{j}$ and $\ket{v_j}$ be the singular values and right singular vectors of $K_U$, with
\begin{equation}
    K_U\ket{v_j} = \sigma_j\ket{g_j}, \qquad
    \sigma_j^2=q\lambda_j \: ,
    \qquad
    |\lambda_j-1|\leq\eta \: .
\end{equation}
Writing
\begin{equation}
    \theta_j \equiv \arcsin\sigma_j \: ,
    \qquad
    \theta_0 \equiv \arcsin\sqrt q \: ,
\end{equation}
one finds that
\begin{equation}
    M_{U}^t UJ\ket{v_j}
    =
    \sin(L_{t} \theta_j) \: 
        \ket{0^m}_P\ket{g_j}_O
    +
    \cos(L_{t} \theta_j) \:
        \ket{b_j} \: , \qquad L_{t} \equiv 2t + 1 \: ,
\end{equation}
where $\ket{b_j}$ lies in the orthogonal complement of the
post-selected subspace. 

Choose $t$ so that $L_{t}$ is the odd integer nearest to
$\frac{\pi}{2\theta_0}$, so
\begin{equation}
    | L_{t} \theta_{0} - \frac{\pi}{2} | \leq \theta_{0} \: .
\end{equation}
Then taking $Q_{U} \equiv M_{U}^{t} U$, one has
\begin{equation}
    \| Q_{U} J | v_{j} \rangle - |0^{m} \rangle | g_{j} \rangle \|_{2} = \sqrt{2 - 2 \sin (L_{t} \theta_{j})}
\end{equation}
It follows that we may take
\begin{equation}
    \| Q_{U} J - | 0^{m} \rangle _{P} \otimes V_{U} \| \leq \max_{j} \sqrt{2 - 2 \sin (L_{t} \theta_{j})} = 2 \max_{j} \Big| \sin(\frac{L_{t} \theta_{j} - \pi/2}{2}) \Big| \: .
\end{equation}

To get a more explicit upper bound, we can use that $\lambda_{j}$ are the normalized eigenvalues of $T_{U}^{2}$ and thus satisfy
\begin{equation}
    | \sqrt{\lambda_{j}} - 1 | \leq \Delta_{\eta} \: ,
\end{equation}
so the mean value theorem gives
\begin{equation}
    | \theta_{j} - \theta_{0} | = | \sin^{-1}(\sqrt{ q \lambda_{j}}) - \sin^{-1}(\sqrt{q}) | = \frac{| \sqrt{q \lambda_{j}} - \sqrt{q}|}{\sqrt{1 - q(1 + \eta)}} \leq \frac{\sqrt{q} \Delta_{\eta}}{\sqrt{1 - q(1 + \eta)}} \: .
\end{equation}
We thus have
\begin{equation}
    \begin{split}
        2 \max_{j} \Big| \sin(\frac{L_{t} \theta_{j} - \pi/2}{2}) \Big| & \leq \max_{j} \Big| L_{t} \theta_{j} - \frac{\pi}{2} \Big| \leq \max_{j} \left[ L_{t} |\theta_{j} - \theta_{0}| + \Big| L_{t} \theta_{0} - \frac{\pi}{2} \Big| \right] \\
        & \leq \theta_{0} + \left( \frac{\pi}{2 \theta_{0}} + 1 \right) \frac{\sqrt{q} \Delta_{\eta}}{\sqrt{1 - q(1 + \eta)}} \\
        & \leq \sin^{-1} \sqrt{q} + \frac{\sqrt{q} + \pi/2}{\sqrt{1 - q(1 + \eta)}} \Delta_{\eta} \: ,
    \end{split}
\end{equation}
using $\sin^{-1} \sqrt{q} \geq \sqrt{q}$, 
which gives the claimed value of $\alpha_{q, \eta}$. 
To get the precise upper bound on $\text{err}(U, \mathcal{D}_{U})$ given in the statement of the theorem, we may simply note that identities
\begin{equation}
    \sin^{-1} \sqrt{q} \leq \frac{\pi}{2} \sqrt{q} \: , \quad \sqrt{1 - q (1 + \eta)} \geq \frac{1}{2} \: , \quad \sqrt{q} \leq \frac{1}{\sqrt{2}} \: , \quad \Delta_{\eta} < (2 - \sqrt{2}) \eta 
\end{equation}
for our regime of parameters $0 < \eta < \frac{1}{2}$ imply
\begin{equation}
    \begin{split}
        2 \Delta_{\eta} + \alpha_{q, \eta} & = 2 \Delta_{\eta} + \sin^{-1}\sqrt{q} + \frac{\sqrt{q} + \pi/2}{\sqrt{1 - q(1+ \eta)}} \Delta_{\eta} \\
        & \leq \frac{\pi}{2} \sqrt{q} + (\pi + \frac{1}{\sqrt{2}} + 2) (2 - \sqrt{2}) \eta \leq  \frac{\pi}{2} (2^{-m/2} + c \eta) \: ,
    \end{split}
\end{equation}
with $c \approx 2.18$. 

Each Grover iteration uses $O(1)$ applications of $U$ and
$U^\dagger$, together with reflections about the subspaces in which the $A$ or $P$ registers are in the zero state.  These
reflections require only $\textnormal{poly}(n)$ elementary gates. Since $t=O(2^{m/2})$, the stated circuit complexity bound follows.
\end{proof}

\section{Efficient decoder for post-selected stabilizer maps}
\label{app:Clifford}

In this appendix, following the discussion of decoders for stabilizer isometries mentioned in section \ref{sec:admi}, we first isolate the stabilizer property responsible for efficient decoding,
and then apply it to uniformly random Cliffords and random stabilizer tensor
networks.

\begin{proposition}[Post-selected stabilizer maps]
\label{prop:postselected-stabilizer}
Let $C$ be an $n$-qubit Clifford unitary and define
\begin{equation}
 K_C=(\langle 0^m|_P\otimes\mathds{1}_O)
 C(\mathds{1}_I\otimes|0^a\rangle_A) \: .
\end{equation}
Either $K_C=0$, or there is a stabilizer-code projector $\Pi_S$ and an integer
$0\leq r\leq m$ such that
\begin{equation}
                         K_C^\dagger K_C=2^{-r}\Pi_S \: .
\label{eq:stabilizer-map-normal-form}
\end{equation}
In particular, if $K_C$ has full column rank, then $\Pi_S=\mathds{1}$ and
$2^{r/2}K_C$ is a stabilizer isometry.  Given a Clifford-circuit description of
$C$, the inverse of $K_{C}$ on the code subspace can be implemented by a Clifford circuit
of size $O(n^2)$.
\end{proposition}

\begin{proof}
Let $\mathcal Z_P$ be the group generated by the $Z$ operators on $P$, and set
\begin{equation}
                         \mathcal G_C=C^\dagger\mathcal Z_P C \: .
\end{equation}
Because $C$ is Clifford, $\mathcal G_C$ is an abelian Pauli subgroup.  Expanding
the post-selection projector gives
\begin{equation}
 K_C^\dagger K_C
 =2^{-m}\sum_{g\in\mathcal G_C}
 (\mathds{1}_I\otimes\langle0^a|_A)
 g(\mathds{1}_I\otimes|0^a\rangle_A) \: .
\label{eq:KC-stabilizer-sum}
\end{equation}
Any term for which the restriction $g_A$ contains an $X$ or $Y$ vanishes.
The remaining elements form a subgroup $\mathcal T\subseteq\mathcal G_C$; sandwiching these with $|0^a\rangle_A$ defines a homomorphism $\rho$ from $\mathcal{T}$ to the signed Pauli group on $I$. Let $N \equiv \text{ker}(\rho)$ and $S \equiv \text{im}(\rho)$. Given that each element of $S$ has $|N|$ pre-images, we obtain
\begin{equation}
K_{C}^{\dagger} K_{C} = 2^{-m} |N| \sum_{s \in S} s \: .
\end{equation}
If $-\mathds{1}_{I} \in S$, the sum vanishes and $K_{C} = 0$. Otherwise, $S$ is a stabilizer group, and $\Pi_{S} \equiv |S|^{-1} \sum_{s \in S} s$ is a projector onto its common $+1$ eigenspace. It follows that $K_{C}^{\dagger} K_{C} = 2^{-r} \Pi_{S}$ where $r = m - \log_{2} | \mathcal{T} |$. 

If $K_C$ has full column rank, the projector $\Pi_S$ must have full rank and is
therefore the identity.  The normalized map $2^{r/2} K_{C}$ is then an isometry obtained from
stabilizer preparations, a Clifford circuit, and stabilizer post-selection. 
Since this map is a stabilizer isometry, stabilizer-code synthesis gives a Clifford unitary $E_{C}$ satisfying
\begin{equation}
    E_{C} (|\psi \rangle_{I} \otimes |0^{a-m} \rangle) = 2^{r/2} K_{C} | \psi \rangle_{I}
\end{equation}
for all logical $| \psi \rangle$. The construction of Cleve-Gottesman \cite{Cleve:1996pp} gives such an encoding circuit using $O(n^{2})$ gates; its inverse is the desired decoder. 
\end{proof}

The integer $r$ in proposition~\ref{prop:postselected-stabilizer} is the number
of independent 
stabilizer constraints imposed by the
post-selection.  It need not equal the literal number $m$ of post-selected
qubits.  The following result shows that $r=m$ with high probability for the
uniform Clifford ensemble.

\begin{theorem}[Efficient decoding for uniformly random Cliffords]
\label{thm:random-clifford}
Let $U$ be uniformly distributed over the $n$-qubit Clifford group.  Then
\begin{equation}
 \Pr_U\!\left[2^mK_U^\dagger K_U=\mathds{1}\right]>1-2^{m-a} \: .
\label{eq:random-clifford-admissibility}
\end{equation}
Consequently the uniform Clifford ensemble is
$(a,m,0,2^{m-a})$-admissible, and on the event in
\eqref{eq:random-clifford-admissibility} the normalized map
$W_U=2^{m/2}K_U$ admits a Clifford decoder of size $O(n^2)$.
\end{theorem}

\begin{proof}
We recall that the $n$-qubit Pauli group (modulo phases) can be identified with the vector space $\mathbb F_2^{2n}$, via the map
\begin{equation}
    X^{x_{1}}Z^{z_{1}} \otimes \ldots \otimes X^{x_{n}} Z^{z_{n}} \rightarrow (x_{1}, \ldots, x_{n} | z_{1}, \ldots, z_{n}) \: .
\end{equation}
See e.g. \cite{Gottesman1997} for an introduction to stabilizer quantum error correction. 
Let
\begin{equation}
 \mathcal G_U=U^\dagger\mathcal Z_PU 
\end{equation}
and let $\mathcal T\subset\mathbb F_2^{2n}$ be the set of Paulis whose
restriction to $A$ contains only $\mathds{1}$ and $Z$.  The Pauli on $I$ is
unrestricted, so
\begin{equation}
                              |\mathcal T|=4^{n-a}2^a=2^{2n-a} \: .
\end{equation}
If $g\notin\mathcal T$, then the restriction $g_{A}$ contains an $X$ or $Y$ and
$\langle0^a|g_A|0^a\rangle=0$.  Consequently, a trivial intersection $ \mathcal G_U\cap\mathcal T=\{\mathds{1}\}$ would imply that
\begin{equation}
    K_U^\dagger K_U=2^{-m}\mathds{1} \: .
\end{equation}

Clifford conjugation acts transitively on the non-identity Pauli operators, so for $U$ uniformly distributed over Cliffords, one has $U g U^{\dagger}$ uniformly distributed over Paulis for fixed Pauli $g$. Thus,
\begin{equation}
                         \Pr_U[g\in\mathcal G_U]=\frac{2^m-1}{4^n-1} \: .
\end{equation}
A union bound over the $2^{2n-a}-1$ non-identity elements of $\mathcal T$ therefore gives
\begin{equation}
 \Pr_U[\mathcal G_U\cap\mathcal T\neq\{\mathds{1}\}]
 \leq (2^{2n-a}-1)\frac{2^m-1}{4^n-1}<2^{m-a}.
\end{equation}
The decoder statement now follows from
proposition~\ref{prop:postselected-stabilizer}.
\end{proof}

The same efficient-decoding mechanism applies to stabilizer tensor networks.
Each vertex stabilizer state has a Clifford preparation circuit, and an
internal tensor-network contraction is a projection onto a maximally entangled
stabilizer state.  After a Clifford change of basis, these contractions are
computational-basis post-selections, so the entire bulk-to-boundary map is a
post-selected stabilizer map.  Whenever it is proportional to an isometry,
proposition~\ref{prop:postselected-stabilizer} supplies a polynomial-size
decoder.

There are natural sufficient conditions for this isometry assumption.  If the
network can be oriented from bulk to boundary so that the incoming legs at
each vertex constitute at most half of its legs, then a perfect tensor at each
vertex acts isometrically from its incoming to outgoing legs, and their
composition is a bulk-to-boundary isometry.  For a random $t$-leg stabilizer
tensor whose legs are prime $p$-dimensional qudits,
\cite{Apel:2021tnn} prove, at large $p$, that the probability that the tensor is perfect is
\begin{equation}
 p_{\text{perfect}}
 \geq
 \max\!\left\{0,
 1-\frac{1}{2p^b}{t\choose\lfloor t/2\rfloor}
 \right\} \: ,
 \qquad 0<b\leq1 \: .
\end{equation}
A union bound therefore makes every tensor perfect,
and hence the whole network isometric, with high probability when the bond
dimension grows sufficiently quickly with the number of vertices.

\section{Proofs of lemmas for query lower bound}
\label{app:lemmas}

In this appendix, we will provide a proof for lemma \ref{lem:frame-extension}, an intermediate result required in the proof of our query lower bound in section \ref{sec:lower-bound}. As a pre-requisite, we will first establish the following simple lemma. 

\begin{lemma} \label{lemma:simple_lemma}
    For any two unit vectors \(x,y\), there exists a
unitary \(R\), acting trivially outside \(\operatorname{span}\{x,y\}\),
such that
\[
    Rx=y
    \qquad\text{and}\qquad
    \|R- \mathds{1}_{d} \|_{\text{op}} = \|x-y\|_{\text{op}} \: .
\]
\end{lemma}

\begin{proof}
    Let $u$ denote the unit normalized component of $y$ orthogonal to $x$, writing\footnote{If $y=\alpha x$ for some $|\alpha|=1$, take
    $R=\mathds{1}_d+(\alpha-1)xx^\dagger$.
    Then $Rx=y$, $R$ is unitary and acts trivially outside the span of $x$, and
    $\|R-\mathds{1}_d\|=|\alpha-1|=\|x-y\|_2$.}
\[
    y=\alpha x+\beta u \: , \qquad \beta \geq 0 \: , \qquad | \alpha|^{2} + \beta^{2} = 1 \: .
\]
With respect to the orthonormal basis $(x, u)$, define $R$ to act on
\(\operatorname{span}\{x,u\}\) as
\[
    R=
    \begin{pmatrix}
        \alpha&-\beta\\
        \beta&\overline{\alpha}
    \end{pmatrix} \: .
\]
This unitary maps \(x\) to \(y\) by construction, and its eigenvalues, which must satisfy
\begin{equation}
    \lambda_{+} \lambda_{-} = | \alpha |^{2} + \beta^{2} = 1 \: , \qquad \lambda_{+} + \lambda_{-} = 2 \text{Re}(\alpha) \: ,
\end{equation}
are given by
\[
    \lambda_{\pm} \equiv e^{\pm i\theta} \: , \qquad \cos\theta \equiv \text{Re}(\alpha) \: .
\]
Thus,
\[
    \|R- \mathds{1}_{d} \|_{\text{op}}
    =
    \sqrt{2-2 \text{Re}(\alpha)}
    =
    \|x-y\|.
\]
\end{proof}

We can now furnish a proof of lemma \ref{lem:frame-extension}, which we recall here for convenience.

\begin{mylemma}{1}
Let \(C,C':\mathbb{C}^2\to\mathbb{C}^d\) be isometries.  Then there exists a
unitary \(L\in U(d)\) such that
\[
    LC=C'
\]
and
\[
    \|L- \mathds{1}_{d} \|_{\text{op}} \leq 3\|C-C'\|_{\text{op}} \: .
\]
\end{mylemma}

\begin{proof}
We may generally write
\begin{equation}
    C= ( c_0, c_1) \: ,
    \qquad
    C'= ( c'_0, c'_1) \: ,
\end{equation}
for some $c_{0}, c_{1}, c_{0}', c_{1}' \in \mathbb{C}^{d}$. We denote
\begin{equation}
    \delta \equiv \|C-C'\|_{\text{op}} \: .
\end{equation}
Then in particular 
\begin{equation}
    \|c_j-c'_j\| \leq \delta \: ,
    \qquad j \in \{ 0,1 \} \: .
\end{equation}

Applying 
the construction of lemma \ref{lemma:simple_lemma} to \(c_0,c'_0\),  we obtain a unitary \(R_0\)
such that
\begin{equation}
    R_0c_0=c'_0 \: ,
    \qquad
    \|R_0- \mathds{1}_{d} \|\leq\delta \: .
\end{equation}
Additionally, let
\begin{equation}
    x \equiv R_0c_1 \: ,
    \qquad
    y \equiv c'_1 \: .
\end{equation}
Both \(x\) and \(y\) are orthogonal to \(c'_0\), and
\begin{equation}
    \|x-y\|
    \leq
    \|(R_0- \mathds{1}_{d} )c_1\|+\|c_1-c'_1\|
    \leq 2\delta \: .
\end{equation}
Applying the construction again, we may therefore choose a unitary \(R_1\), acting inside
\((c'_0)^\perp\) and fixing \(c'_0\), such that
\begin{equation}
    R_1x=y \: ,
    \qquad
    \|R_1-\mathds{1}_{d} \|\leq 2\delta \: .
\end{equation}
Finally, let 
\begin{equation}
    L \equiv R_1R_0 \: .
\end{equation}
Then \(LC=C'\), and
\begin{equation}
    \|L- \mathds{1}_{d} \|_{\text{op}}
    \leq
    \lVert R_{1} \rVert_{\text{op}} \|R_0- \mathds{1}_{d}\|_{\text{op}}+\|R_1-\mathds{1}_{d} \|_{\text{op}}
    \leq 3\delta \: ,
\end{equation}
as desired. 
\end{proof}

\section{Upper bound for structured Haar tensor network}
\label{app:learningbound}
The setting we have in mind is a tensor network made of independent Haar tensors, with a known structure. The query model gives access to each Haar tensor individually. We will prove an upper bound on the query complexity needed to solve an analogous one-bit decoding task using learning methods. To show this, we first construct the appropriate learning theorem from the results of \cite{haah2023query}:
\begin{theorem}\label{thm:learning}
    There is a quantum algorithm $\mathcal{A}$ that, given black-box access to an unknown $d$-dimensional unitary $U\in U(d)$ and any $\delta>0$ and $\eta>0$, outputs a classical description of a unitary $\hat{U}$ such that
    \begin{enumerate}
        \item It uses $O\left(\frac{d^2}{\delta}\log\left(\frac{1}{\eta}\right)\right)$ queries.
        \item It satisfies
        \begin{align}\label{eq:learning_probability}
            \Pr[d_{ph}(U,\hat{U})>\delta)]\leq\eta
        \end{align}
        for
        \begin{align}\label{eq:dph_defn}
            d_{ph}(U,V) := \min_{|\omega|=1} \|U-\omega V\|_{op}.
        \end{align}
    \end{enumerate}
\end{theorem}
\begin{proof}
    This theorem combines a series of statements from \cite{haah2023query}. From Theorem 1.1 in \cite{haah2023query} we have an algorithm $\mathcal{A}$ that uses at most $O(d^2/\epsilon)$ queries and produces a classical description of $U$ within a constant probability in diamond norm:
    \begin{align}
        \Pr[\|\mathcal{U}(U)-\mathcal{U}(\hat{U})\|_\diamond\leq 3\epsilon] \geq \frac{2}{3},
    \end{align}
    where $\mathcal{U}(\cdot)$ notates the corresponding unitary channel.
    Proposition 2.4 from \cite{haah2023query} then provides a method of confidence boosting by calling the algorithm $\mathcal{A}$ $O(\log(1/\eta))$ times to guarantee that one of the classical descriptions generated satisfies:
    \begin{align}
        \Pr[\|\mathcal{U}(U)-\mathcal{U}(\hat{U})\|_\diamond\leq 6\epsilon] \geq \eta.
    \end{align}
    Lastly, from Proposition 1.6 in \cite{haah2023query}, we know
    \begin{align}
        d_{ph}(U,V)\leq \|\mathcal{U}(U)-\mathcal{U}(V)\|_\diamond.
    \end{align}
    We take $\delta=6\epsilon$ to complete the proof.
\end{proof}
We also prove the following technical lemma about Haar unitaries since it will be useful for the proof; in essence, it says that the Haar tensor network is unlikely to produce an output state that has too small a norm.
\begin{lemma}\label{lem:haarball}
    Let $A:\mathbb{C}^{d^2}\rightarrow \mathcal{H}$ be a fixed nonzero linear map and let $U$ be Haar in $U(d)$. Then for every $t>0$,
    \begin{align}\label{eq:haar_ball_probability}
        Pr\left[\|A \,\text{vec} (U)\|\leq t\|A\|_{op}\right] \leq \min\{1,2dt\}.
    \end{align}
\end{lemma}
\begin{proof}
    Let $\sigma=\|A\|_{op}$ and consider $w$, the first left singular vector of $A$ with norm 1. By definition, this means
    \begin{align}
        AA^\dagger w = \sigma^2 w,
    \end{align}
    or in other words there exists a vector $v$ such that 
    \begin{align}
        A^\dagger w = \sigma v.
    \end{align}
    Consequently, for any vector $x$,
    \begin{align}
        w^\dagger A x = \sigma v^\dagger x.
    \end{align}
    Now we take $x=\text{vec}(U)$. Recall that the dot product of two vectors is equivalent to the trace of multiplying their matrix counterparts.
    Defining $B$ as the unique matrix such that $\text{vec} (B)=v$, we find $B$ satisfies 
    \begin{align}\label{eq:relate_B}
        \omega^\dagger A\, \text{vec} (U) = \sigma \tr(B^\dagger U) \implies \| A \, \text{vec} (U)  \|= \|A\|_{op} \tr(B^\dagger U).
    \end{align}
    As such we can now focus on the probability of the event $\tr(B^\dagger U)\leq t$.
    Notice 
    \begin{align}
        \|B\|_F = \|v\| = \left(\frac{A^\dagger w }{\sigma}\right)\left(\frac{A^\dagger w }{\sigma}\right)^\dagger = \|w\|=1.
    \end{align}
    As such, there must exist a row $a$ in $B$ such that $\|B_{a,*}\|\geq 1/\sqrt{d}$. We now analyze this row in a clever way to bound probabilities on the whole matrix. Let $\theta$ be a uniformly random phase variable, and let $V$ be a Haar random unitary. Because of left invariance of Haar, we can rewrite Haar $U$ as $U=E_a(\theta)V$ where $E_a(\theta)$ solely multiplies the row $a$ by $e^{i\theta}$. Then
    \begin{align} \label{eq:defn_bc}
        \tr(B^\dagger U) = b + e^{i\theta} c,\quad c=\sum_{j=1}^d B^\dagger_{aj}V_{aj}.
    \end{align}
    Notice $b$, $c$ are just complex numbers that depend on $v$ and $V$. Given the uniform randomness of $\theta$, we can then ask the likelihood that $(b+e^{i\theta}c)<t$ for fixed $b$ and $c$. First consider $t<|c|$. Geometrically, this is equivalent to asking how much of the circle drawn by varying $\theta$ in $(b+e^{i\theta}c)$ is within the circle of radius $t$ around the origin. The overlapping arc of the circle has endpoints at $\theta_1$ and $\theta_2$ such that
    \begin{align}
        |b+e^{i\theta_1}c|<t,\quad|b+e^{i\theta_1}c|<t.
    \end{align}
    By triangle inequality, this implies the distance between these two endpoints is less than $2t$. Given that the chord length of an angle $\Delta\theta$ is given by $2r\sin(\Delta\theta/2)$, we find
    \begin{align}
        2r\sin(\Delta\theta/2)\leq 2t\implies\Delta\theta\leq 2\arcsin(t/r) < 2\pi t/r.
    \end{align}
    Then, the arclength $\Delta\theta/2$ gives the probability bound for $t<c$, while 1 bounds $t\geq c$:
    \begin{align}\label{eq:pr_c}
        \Pr_\theta[|b+e^{i\theta}c|\leq t]\leq \min\left\{1,\frac{\Delta\theta}{2\pi}\right\}\leq\min\left\{1,\frac{t}{|c|}\right\}.
    \end{align}
    Now we look at $b$ and $c$. Notice the bound on the righthand side only depends on $c$, so we focus on controlling an average on $c$ through the Haar distribution of $V$. Since $B$ is fixed by fixing $A$, the distribution of $c$ is entirely controlled by the distribution of the entries of $V_{aj}$, which are independent complex Gaussians appropriately normalized to give a unit vector. Due to the rotational invariance of $V_{aj}$, the overlap computed by $c$ is equivalent to the overlap of $V_{aj}$ with any unit vector, or in other words,
    \begin{align}
        \frac{c^2}{\|B_{a,*}\|^2}\stackrel{\text{dist.}}{\sim} V_{a,1}.
    \end{align}
    This normalized Gaussian vector entry is known to follow the distribution of $\text{Beta}(1,d-1)$ (see Appendix G of \cite{ghosh2026deepthermalizationquantumchaos}), which obeys the probability
    \begin{align}
        \Pr[V_{a,1}<y] = 1-(1-y)^{d-1}.
    \end{align}
    By Bernoulli's inequality, we then find
    \begin{align}
        \Pr_c[|c|\leq u] \leq \min\left\{1,\frac{(d-1)u^2}{\|B_{a,*}\|^2}\right\}.
    \end{align}
    Then we can integrate the appropriate tail bound to find the average over $c$:
    \begin{align}
        \mathbb{E}_c \frac{1}{|c|} &= \int_0^\infty \Pr[|c|< x^{-1}]dx \\
        &= \frac{\sqrt{d-1}}{\|B_{a,*}\|} +\int_{\frac{\sqrt{d-1}}{\|B_{a,*}\|}}^\infty \left(\frac{\sqrt{d-1}}{\|B_{a,*}\|}\right)^2 \frac{1}{x^2}dx \\&= 2* \frac{\sqrt{d-1}}{\|B_{a,*}\|} \leq 2d
    \end{align}
    Combining this bound with \eqref{eq:pr_c}, \eqref{eq:defn_bc}, and \eqref{eq:relate_B}, and averaging, completes the proof.
\end{proof}
Finally we introduce and prove an upper bound on the query complexity for the one-bit decoding task for a structured Haar tensor network.
\begin{theorem}
    Suppose \(1\leq m<n\) and let $d = 2^{n}$.  
    Let $U_1,...,U_N\in U(d)$ be a set of $N$ independently drawn Haar random unitaries sitting in a known (a,m,$\eta_{adm}$,$\delta_{adm}$)-admissible tensor network structure $K(\{U\}_i)$ with $m$ post-selections, and let $k$ be uniformly distributed in $\{0, 1\}$ independently of any $U_i$. Let
    \begin{equation}
        f_k(\{U\}_i) = K(\{U\}_i)\ket0^{\otimes n-1}\ket k,\quad \ket{\Psi_k(K(\{U\}_i))} = \frac{f_k(\{U\}_i)}{\|f_k(\{U\}_i)\|},
    \end{equation}
    such that $\ket{\Psi_k}$ is well-defined for both $k$.
    Then given one copy of \(\ket{\Psi_k(K(\{U\}_i))}\) for unknown $k$, there exists a quantum algorithm  \(\mathcal A\) that can recover $k$,
    with at most 
    \begin{equation}\label{eq:query_number}
        T = O\left(\frac{N^3d^{7/2}}{\varepsilon\beta}\log\left(\frac{2N}{\beta}\right)\right)
    \end{equation}
    queries to \(U\), and average success probability
    \begin{equation}
    \mathbb{E}[p_{\text{succ}}(U, \mathcal{A})] \geq p_{opt}-2\epsilon-\beta,\quad p_{opt}= \frac{1}{2}+\frac{1-\delta_{adm}}{2}\sqrt{1-\eta^2_{adm}}.
    \end{equation}
\end{theorem}
\begin{proof}
    We begin by applying the learning protocol from \ref{thm:learning} on each individual Haar random tensor in the network with the following parameters:
    \begin{align}
        \eta = \frac{\beta}{2N},\quad\delta = \frac{\varepsilon\beta}{16N^2D^{3/2}}
    \end{align}
    Altogether, this accounts for every query in the query upper bound in \eqref{eq:query_number}. The decoder then works as following: using the learned tensors $\{\hat{U}\}_i$, the decoder classically contracts the tensor network to find the states $\hat{\rho}_1$ and $\hat{\rho}_0$ corresponding to different inputs $k=1$ and $k=0$. The decoder then classically determines the measurement $\hat{M}$ that optimally distinguishes $\hat{\rho}_1$ and $\hat{\rho}_0$, and then applies it to the received state $\ket{\Psi_k(K(\{U\}_i))}$ to find $k$. 
    
    The rest of this proof is dedicated to analyzing the success probability of this protocol. theorem \ref{thm:learning} gives us the errors introduced by each learned tensor; we will now propagate these through the tensor network to see how this affects the error of $\|\hat{\rho}_k-\rho_k\|_1$. We set up some notation. We define an ordered set of unitaries in the tensor network $K$ where the first $j$ are replaced with the learned unitaries and the last $N-j$ are still the original:
    \begin{align}
        U^{(j)} = \{\hat{U}_1,...,\hat{U}_j,U_{j+1},...,U_N\},\quad j=0,...,N.
    \end{align}
    Thanks to the multilinearity of the tensor network $K$, we can rewrite the output $f$ as
    \begin{align}
        f_k(U^{(j-1)}) = A_{j,k}\text{vec}(U_j),\quad f_k(U^{(j)}) = A_{j,k}\text{vec}(\hat{U_j}).
    \end{align}
    The error in $\|\hat{\rho}_k-\rho_k\|_1$ will be the sum of errors from each step of replacing one tensor at a time; each step's error can be calculated by the difference of these two outputs.

    A minor technical note: we will actually take the $\hat{U}_j$ here not to be the actual learned tensors $\hat{U}_j$ but rather $\omega_j\hat{U}_j$ where $\omega$ is defined as the $\omega$ realizing the minimum from \eqref{eq:dph_defn}. This is for ease of dealing with the operator norm, and because thanks to the homogeneity of the tensor network $K$, the $\hat{\rho}_k$ generated by $\{\hat{U}\}_i$ and $\{\omega_i \hat{U}\}_i$ are the same.

    Now we turn to $\|f_k(U^{(j-1)})-f_k(U^{(j)})\|$. We consider the event $G$ where both (1) every tensor is learned well and satisfies
    \begin{align}
        d_{ph}(U_j,\hat{U}_j)>\delta,
    \end{align}
    and (2) for every step $j$, $A_{j,k}$ does not produce a state with too small a norm and satisfies
    \begin{align}
        \|A_{j,k} \text{vec}(U_j)\|\leq t\|A_{j,k}\|_{op}
    \end{align}
    for 
    \begin{align}
        t=\frac{\beta}{8Nd}.
    \end{align}
    Then by \eqref{eq:learning_probability} in theorem \ref{thm:learning} and union bound over all $N$ tensors, the first condition only fails with at most probability $N*\eta=\beta/2$. By \eqref{eq:haar_ball_probability} from lemma \ref{lem:haarball} and the union bound over all $A_{j,k}$, the second condition only fails with probability at most $4dtN=\beta/2$. As such, the probability of the event $G$ is at least $1-\beta$.

    In the event $G$, by condition (1) we find
    \begin{align}
        \|f_k(U^{(j-1)})-f_k(U^{(j)})\|&\leq \|A_{j,}\|_{op}\|U_j-\hat{U}_j\|_F\\
        &\leq \sqrt{d}\|A_{j,}\|_{op}\|U_j-\hat{U}_j\|_{op}\\
        &\leq \delta\sqrt{d}\|A_{j,}\|_{op}.
    \end{align}
    By condition (2), we find
    \begin{align}
        \|f_k(U^{(j-1)})\| \geq t\|A_{j,}\|_{op}.
    \end{align}
    By triangle inequality, we know that for two unnormalized vectors $u$,$v$:
    \begin{align}
        \left\|\frac{v}{\|v\|}-\frac{u}{\|u\|}\right\| = \left\|\frac{v}{\|v\|}-\frac{u}{\|u\|}+\frac{v}{\|u\|}-\frac{v}{\|u\|}\right\|\leq 2\frac{\|u-v\|}{\|u\|}.
    \end{align}
    Then we find
    \begin{align}
        \left\|\frac{f_k(U^{(j-1)})}{\|f_k(U^{(j-1)})\|}-\frac{f_k(U^{(j)})}{\|f_k(U^{(j)})\|}\right\|&\leq \frac{2\sqrt{d}\delta}{t}=\frac{\varepsilon}{N}.
    \end{align}
    Then we can relate this to the states:
    \begin{align}
        \frac{1}{2}\|\rho_k-\hat{\rho}_k\|_1\leq\left\|\sum_N\ket{\Psi^{(j-1)}}-\ket{\Psi^{(j)}}\right\| = \varepsilon
    \end{align}
    We can then turn this into the success probability for event $G$. Let the probability of success of distinguishing states $\rho,\sigma$ given measurement $M$ be denoted $p(M;\rho,\sigma)$. Then from our 1-norm bound, we have
    \begin{align}
        |p(M;\rho_0,\rho_1)-p(M;\hat{\rho}_0,\hat{\rho}_1| = \frac{1}{2}|\tr(M(\rho_0-\hat{\rho}_0) - \tr((\mathds{1} - M)(\rho_1-\hat{\rho}_1) )|\leq \epsilon.
    \end{align}
    For ${M}_{opt}$, $\hat{M}_{opt}$ defined as the optimal measurements for the original and the learned case respectively, we then can compute
    \begin{align}
        p_{succ,\,G} =p(\hat{M}_{opt};\rho_0,\rho_1)&\geq p(\hat{M}_{opt};\hat{\rho}_0,\hat{\rho}_1)-\varepsilon\\
        &\geq p({M}_{opt};\hat{\rho}_0,\hat{\rho}_1)-\varepsilon\\
        &\geq p({M}_{opt};\rho_0,\rho_1)-\varepsilon-\varepsilon\\
        &= p_{opt}-2\varepsilon.
    \end{align}
    Now consider the situation outside event $G$, which occurs with probability at most $\beta$. Compared $p_{succ,G}$, the most the probability can be smaller is by 1:
    \begin{align}
        p_{succ,not\,G} = p_{opt}-2\varepsilon-1.
    \end{align}
    Considering both events and averaging then gives us
    \begin{equation}
    \mathbb{E}[p_{\text{succ}}(U, \mathcal{A})] = p(G)p_{succ,\,G}+p(not\,G)p_{succ,not\,G}\geq p_{opt}-2\varepsilon-\beta.
    \end{equation}
    The last piece of this proof is to find the form of $p_{opt}$. The probability of distinguishing the two states is controlled by the admissibility of the tensor network. In particular, the admissibility condition tells us
    \begin{equation}
        \underset{\{U\}_i \sim \mathcal{E}}{\textnormal{Pr}} \left[ \| 2^{m} K(\{U\}_i)^{\dagger} K(\{U\}_i) - \mathds{1}_{I} \| \leq \eta_{adm} \right] \geq 1 - \delta_{adm} \: .
    \end{equation}
    Let's first consider the event that the admissibility condition is satisfied, which occurs with probability $1-\delta_{adm}$. By the definition of operator norm, the admissibility condition implies for any state (possibly unnormalized) $\ket\psi$,
    \begin{align}
        \bra\psi  2^{m} K(\{U\}_i)^{\dagger} K(\{U\}_i) - \mathds{1}_{I} \ket\psi \leq \eta_{adm}\|\ket\psi\|^2
    \end{align}
    This in turn implies  
    \begin{align}\label{eq:adm_inequality}
        (1-\eta_{adm})\|\ket\psi\|^2 \leq \|2^{m/2}K\ket\psi\|^2\leq (1+\eta_{adm})\|\ket\psi\|^2.
    \end{align}
    Since both $\ket{\Psi_1(\{U\}_i)}$ and $\ket{\Psi_0\{U\}_i}$ are pure, the optimal distinguishability is
    \begin{align}
        p_{opt}(\{U\}_i) = \frac{1}{2}+\frac{1}{4}\|\rho_0-\rho_1\|_1 = \frac{1+\sqrt{1-|\bra{\Psi_1}\Psi_0\rangle|^2}}{2}.
    \end{align} 
    We will now bound $|\bra{\Psi_1}\Psi_0\rangle|$.
    Consider the states 
    \begin{align}
        \ket{\psi_+} = \ket{\psi_0}+\ket{\psi_1},\quad\ket{\psi_-}=\ket{\psi_0}-\ket{\psi_1}, 
    \end{align}
    where $\psi_1$ and $\psi_0$ are defined as the preimage states
    \begin{align}
        2^{m/2}K\ket{\psi_0}=\ket{\Psi_0}, \quad 2^{m/2}K\ket{\psi_1}=\ket{\Psi_1}. 
    \end{align}
    Based on this, we can calculate
    \begin{align}
        \|2^{m/2}K\ket{\psi_+}\|^2 = 2+2|\bra{\Psi_0}\Psi_1\rangle|^2,\quad \|2^{m/2}K\ket{\psi_-}\|^2 = 2-2|\bra{\Psi_0}\Psi_1\rangle|^2
    \end{align}
    Meanwhile, $\|\ket{\psi_+}\|=\|\ket{\psi_-}\|$ because the cross term $|\langle\psi_0\ket{\psi_1}|=0$ via the setup of the inputs on the one-bit coding problem. Then using \eqref{eq:adm_inequality} twice, we have
    \begin{align}
    \frac{2+2|\bra{\Psi_0}\Psi_1\rangle|^2}{1+\eta_{adm}}&\leq \|\ket{\psi_+}\|=\|\ket{\psi_-}\|\leq \frac{2-2|\bra{\Psi_0}\Psi_1\rangle|^2}{1-\eta_{adm}}\\
    &\implies\, |\bra{\Psi_1}\Psi_0\rangle|\leq \eta_{adm}.
    \end{align}
    This means that in the event that the admissibility condition is satisfied, the optimal probability is
    \begin{align}
        p_{opt,adm}\leq \frac{1+\sqrt{1-\eta_{adm}^2}}{2}.
    \end{align}
    In the event that the admissibility condition is not satisfied, the optimal probability is bounded by random guessing, $p_{opt, not\, adm}\geq1/2$, which occurs with probability $\delta_{adm}$. Combining both probabilities, we conclude
    \begin{align}
        \mathbb{E}_{\{U\}_i} p_{opt}(\{U\}_i) = \frac{1}{2} - \frac{1-\delta_{adm}}{2}\sqrt{1-\eta^{2}_{adm}}.
    \end{align}
\end{proof}

\bibliographystyle{JHEP}
\bibliography{refs}

\end{document}